\documentclass[reprint,onecolumn,12pt,superscriptaddress,nofootinbib,floatfix,amsmath,amssymb,aps,longbibliography]{revtex4-2}
\usepackage[T1]{fontenc}
\usepackage[utf8]{inputenc}
\usepackage{graphicx}
\usepackage{bm}
\usepackage{dcolumn}
\usepackage[colorlinks=true,linkcolor=blue,citecolor=blue,urlcolor=blue]{hyperref}
\usepackage{orcidlink}
\graphicspath{{figures/}}
\newcommand{\orcidJuan}{0000-0003-3756-5016}

\begin{document}
\title{Fundamental-derivative criticality of ion-acoustic solitons in fluid-Poisson plasmas with general equations of state}
\author{Hern\'an Pastenes}
\affiliation{Escuela de Qu\'imica, Facultad de Ciencias Naturales, Matemática y del Medio Ambiente, Universidad Tecnológica Metropolitana, Santiago, Chile.}
\author{Juan F. Mar\'in\orcidlink{\orcidJuan}}
\email{j.marinm@utem.cl}
\affiliation{Departamento de Física, Facultad de Ciencias Naturales, Matemática y del Medio Ambiente, Universidad Tecnológica Metropolitana, Santiago, Chile.}

\begin{abstract}
\vspace{0.5cm}
\begin{center}
\textbf{ABSTRACT}
\end{center}
\vspace{-0.5cm}
Non-ideal ion compressibility modifies the nonlinearity of ion-acoustic waves. In a fluid-Poisson plasma with Boltzmann electrons and a barotropic ion equation of state, we relate the quadratic KdV coefficient to the fundamental derivative of the effective acoustic pressure. Its vanishing marks loss of genuine nonlinearity. Higher derivatives of the equation of state determine whether the cubic reduction is focusing or defocusing and govern the quartic-flux reduction at simultaneous criticality. The exact Sagdeev formulation resolves compressive and rarefactive soliton branches and their Gardner unfolding. Exact continuation identifies an equation-of-state-controlled rarefactive double-layer boundary distinct from both the cubic critical point and an orbitwise pressure-slope threshold. As they approach this boundary, solitons develop a logarithmically widening flat top, and the exact Sagdeev quadrature quantitatively captures the standard saddle-passage law. Along the boundary toward the acoustic codimension-two point, exponential localisation lengths diverge while the amplitude vanishes. Full fluid-Poisson simulations confirm finite-time propagation of selected exact stationary profiles. Our results connect acoustic criticality, exact orbit admissibility, and equation-of-state-controlled solitary waves in dispersive plasmas.
\end{abstract}

\maketitle

\section{Introduction}
\label{Sec:Intro}

Nonlinear waves in dispersive media can form coherent structures whose propagation is controlled by a balance between wave steepening and dispersion \cite{Peyrard2004}. This competition appears in a broad range of physical systems, from shallow-water waves and nonlinear optical media to acoustic excitations in plasmas \cite{Peyrard2004, Shukla1983, Tran1979}. In the latter, ion-acoustic waves provide a particularly useful setting because the ion inertia, electron response, and the electrostatic restoring force can be described within a fluid-Poisson (fP) framework \cite{Tran1979, Haruichi1966}. In the weakly nonlinear, long-wavelength regime, the Korteweg-de Vries (KdV) equation governs their evolution, as established by the seminal works of Washimi and Taniuti and others \cite{Haruichi1966, Taniuti1968, JeffreyKakutani1972}, while laboratory experiments have demonstrated the formation of ion-acoustic solitons \cite{IkeziTaylorBaker1970, Ikezi1973}. These localised structures provide experimental evidence that the dispersion characteristics of the medium govern the propagation dynamics of nonlinear excitations. 

The standard KdV description relies on a non-vanishing quadratic nonlinearity, which fixes the polarity, amplitude and propagation velocity of small-amplitude solitons \cite{Haruichi1966, Tran1979}. However, this nonlinear contribution can disappear when a control parameter reaches a critical value. In this scenario, the modified KdV (mKdV) equation captures the leading cubic nonlinearity, whereas the Gardner equation captures the competition between quadratic and cubic terms as one approaches the critical point \cite{NakamuraTsukabayashi1984, VerheestHereman2025}. Critical KdV-mKdV-Gardner transitions in plasmas are well-established phenomena \cite{Watanabe1984IonAcoustic, VerheestHereman2025}, most often accessed by adjusting the composition or response of additional plasma species. For instance, Watanabe \cite{Watanabe1984IonAcoustic} studied a negative-ion plasma, and modified-KdV solitons were subsequently observed at a critical composition \cite{NakamuraTsukabayashi1984}. Verheest \cite{Verheest2015CriticalDensities} investigated critical densities, while Olivier and coworkers \cite{OlivierVerheestMaharaj2016} used near-critical Sagdeev expansions to classify polarity coexistence, weak double layers and acoustic-speed solitons. The simultaneous suppression of quadratic and cubic nonlinearities has likewise been analysed for supercritical plasma compositions \cite{VerheestOlivierHereman2016}. These precedents establish the occurrence of Gardner solitons and near-critical Sagdeev branches. Therefore, a natural question remains: how derivatives of a prescribed ion equation of state (EOS) control the same nonlinear degeneracies and connect them to the acoustic characteristics of the parent fP system?

The propagation of solitons in a plasma under weak non-ideal effects has been increasingly studied in the literature. Existing studies address finite ion temperature \cite{Tagare1975, Nejoh1987, Chatterjee1994, Chatterjee1997}, non-Maxwellian electron distributions \cite{Sultana2022}, ionisation and collisional damping \cite{Ghosh2005, Jyoti2011, Misra2012}, viscosity and dissipative wave dynamics \cite{Borhanian2013, Goswami2023}, Landau damping \cite{Wong1964, Alexeff1967, Ott1969, Mushtaq2024}, and quantum corrections \cite{Sahu2012, Misra2015, Rajabi2023}. Such mechanisms should be distinguished from the density dependence of a local barotropic ion-pressure law, which is the constitutive ingredient of the present work. In non-ideal plasmas, interparticle correlations, excluded-volume effects and other departures from ideal-gas behaviour modify the pressure and its derivatives \cite{Ichimaru1982, Kremp1984, Chabrier1998, Potekhin2009}. Correlation-informed equations of state and sound-speed estimates are also available for screened Yukawa fluids \cite{KhrapakThomas2015EOS, KhrapakThomas2015Sound}. Earlier studies of non-ideal dust-acoustic waves considered van der Waals-type closures for the dust component \cite{Rao1998, Maharaj2007}. More recently, Kitayev and Dubinov investigated ion-acoustic waves with a van der Waals ion EOS \cite{Kitayev2025}. These prior models establish that non-ideal constitutive laws affect nonlinear plasma waves, but their physical closures and questions differ from the EOS-controlled fundamental-derivative and Sagdeev connections. Gardner dynamics, on the other hand, have also been investigated in multicomponent plasmas with nonthermal electron responses \cite{Guo2015,VerheestHereman2025}.

However, a related perspective emerges in non-classical gas dynamics. In ordinary gases, compression generally increases the propagation speed of acoustic disturbances, whereas sufficiently non-ideal gases may exhibit the opposite behaviour. The fundamental derivative, introduced in gas dynamics by Thompson \cite{Thompson1971Fundamental}, characterises how acoustic characteristic speeds depend on density. Its vanishing signals a loss of genuine nonlinearity in the sense of hyperbolic conservation-law theory \cite{Lax1957Hyperbolic,CramerKluwick1984}. Importantly, the relation between genuine nonlinearity, general pressure laws and weakly nonlinear dispersive dynamics has already been developed for dispersive Eulerian fluids \cite{Hoefer2014DispersiveEulerian,ElHoeferShearer2017}. Thus, it is useful to determine whether the vanishing quadratic coefficient has a direct interpretation in terms of the characteristic structure of the parent fluid equations, how the associated soliton branches are organised by the exact travelling-wave problem, and whether these asymptotic structures persist under the evolution of the full fP system. Answering these questions requires separating the general effects of a prescribed EOS from the additional constraints needed to realise the closure in a plasma.

In this work, we investigate the nonlinear acoustic response of a one-dimensional fP plasma with a general barotropic ion EOS. For this model, we derive the quadratic KdV coefficient from the fundamental derivative of an effective pressure law that combines the ion response with the quasineutral electron contribution. Its zero identifies the loss of genuine nonlinearity of the right-propagating acoustic family. On that manifold, the density derivative of the fundamental derivative sets the cubic mKdV coefficient. We then relate these coefficients to the exact local Sagdeev derivatives and compare the resulting EOS-controlled KdV, mKdV and Gardner predictions with the admissible finite-amplitude travelling-wave branches. We establish the explicit correspondence between acoustic, reduced-wave and exact stationary descriptions, and show how the near-critical structures are organised. We also solve the full nonlinear equations numerically and compare the resulting coherent structures with the predictions of the reduced models, including systematic finite-amplitude departures from leading-order theory. For a cubic EOS, we additionally determine the exact double-layer locus and a distinct orbitwise pressure-slope boundary, using the standard logarithmic saddle-passage law as a quantitative signature on the flat-top profiles. The EOS-dependent double-layer locus connects to the higher-order critical point, while an exact acoustic-limit argument excludes a surviving finite-amplitude wave there. We treat the EOS as a prescribed local closure and examine the physical realisability of specific critical parameter sets separately from the general mathematical mechanism.

The remainder of the article is organised as follows. Section~\ref{sec:model} introduces the fP model, the non-ideal ion-pressure closure, and the numerical formulation. In Section~\ref{sec:linear}, we examine the linear acoustic response and establish its connection with the effective sound speed and the fundamental derivative. Section~\ref{sec:nonlinear} develops the weakly nonlinear reductions leading to the KdV, mKdV and Gardner equations, identifying the conditions under which the quadratic and cubic nonlinearities become critical. Section~\ref{sec:sagdeev} derives the exact Sagdeev travelling-wave formulation, providing the basis for the soliton bifurcation analysis presented in Section~\ref{sec:bifurcations}. There, we examine how variations in the EOS reorganise the admissible soliton branches, the exact double-layer boundary and its flat-top limit, and the associated spatial dynamics, supported by direct numerical simulations of selected solitons in the full fP system. Section~\ref{sec:eos_applicability} discusses the physical applicability and limitations of the constitutive description, while Section~\ref{sec:conclusions} summarises our main findings.

\section{Fluid-Poisson model with a non-ideal ion equation of state}
\label{sec:model}

\subsection{The governing equations}
\label{sec:model_governing}

Consider a one-dimensional, unmagnetised plasma composed of singly charged positive ions and electrons. The ions have mass $m_i$ and reference temperature $T_i$, whereas the much lighter electrons are assumed to respond on a faster time scale and to remain in local Boltzmann equilibrium at temperature $T_e$. We neglect collisions, viscosity and magnetic effects, so that dispersion arises from the finite electrostatic screening length rather than an imposed higher-order derivative in the fluid equations. The Boltzmann closure assumes a sufficiently classical, weakly coupled electron population. The ions are described by a compressible barotropic fluid with pressure $P_i(N_i)$ along a specified thermodynamic path. This closure offers a controlled framework for isolating the effects of non-ideal compressibility.

We normalise the ion number density $N_i$ and electron density $N_e$ by the common equilibrium density $N_0$, the electrostatic potential $\Phi_{\text{dim}}$ by $k_{\mathrm B}T_e/e$, and the ion velocity by $c_s=\sqrt{k_{\mathrm B}T_e/m_i}$, where $k_{\mathrm B}$ is the Boltzmann constant. Space is measured in units of the electron Debye length $\lambda_D=\sqrt{\varepsilon_0 k_{\mathrm B}T_e/(N_0e^2)}$, and time in units of the inverse ion plasma frequency $\omega_{pi}^{-1}=\sqrt{\varepsilon_0m_i/(N_0e^2)}$, with $e$ the elementary charge and $\varepsilon_0$ the vacuum permittivity. We denote the corresponding dimensionless variables by $n:=N_i/N_0$, $n_e:=N_e/N_0$, and $\phi:=e\Phi_{\text{dim}}/(k_{\mathrm B}T_e)$. The ion pressure is normalised as $p(n):=P_i(N_0n)/(N_0 k_{\mathrm B}T_i)$, and $\sigma:=T_i/T_e$ denotes the reference ion-to-electron temperature ratio. Thus, the resulting fP system, written in terms of the dimensionless speed $u$, space coordinate $x$ and time $t$, becomes
\begin{subequations}
\label{eq:fp}
\begin{align}
\partial_tn+\partial_x(nu)&=0,\label{eq:fp_continuity}\\
\partial_tu+u\partial_x u&=-\partial_x\phi-\frac{\sigma}{n}\partial_x p(n),\label{eq:fp_momentum}\\
\partial_{xx}\phi&=e^\phi-n.\label{eq:fp_poisson}
\end{align}
\end{subequations}
Here, $n_e=\exp\phi$ follows from the Boltzmann-electron assumption. The pressure gradient accounts for the prescribed local ion response, while the electric force and Poisson equation retain the self-consistent electrostatic coupling. The homogeneous reference state corresponds to $n=n_e=1$, $u=0$ and $\phi=0$.

\subsection{Non-ideal pressure closure}
\label{sec:model_eos}

We consider the dependence of ion pressure on density. For an isothermal ion closure, it is possible to write $p(n)=nZ(n)$, where $Z(n)$ is the dimensionless compressibility factor. More generally, the barotropic function $p(n)$ can represent other prescribed thermodynamic paths, provided the corresponding derivatives are evaluated consistently along that path. We assume that $p(n)$ is sufficiently smooth near the reference state and expand it in a Taylor series as follows,
\begin{equation}
\label{eq:pressure_expansion}
p(n)=p_0+p_1(n-1)+p_2(n-1)^2+p_3(n-1)^3+\mathcal{O}\bigl([n-1]^4\bigr),
\end{equation}
where $p_0:=p(1)$, $p_1:=p'(1)$, $p_2:=p''(1)/2$ and $p_3:=p'''(1)/6$, with primes denoting differentiation with respect to $n$.

Since only the pressure gradient enters the momentum equation, the constant $p_0$ has no dynamical effect and may be set to zero in numerical simulations without implying that the equilibrium pressure vanishes. As we will see later in this article, the first derivative $p_1$ modifies the linear acoustic response, while the curvature $p_2$ and the next coefficient $p_3$ determine the leading nonlinear corrections. Importantly, specifying these local coefficients does not, by itself, establish a microscopic EOS for a particular strongly coupled plasma. For a correlation-based closure, the local ion pressure must be distinguished from the mean-field electron response already represented by Poisson's equation, so we do not double-count electrostatic contributions.

\subsection{Numerical formulation}
\label{sec:model_numerics}

To compare the weakly nonlinear predictions with the parent system, we have numerically integrated Eqs.~\eqref{eq:fp} directly. Introducing the ion momentum density $m:=nu$, the evolution equations take the following conservative form,
\begin{subequations}
\label{eq:fp_conservative}
\begin{align}
\partial_t n&=-\partial_x m,\\
\partial_t m&=-\partial_x\left[\frac{m^2}{n}+\sigma p(n)\right]-n\partial_x\phi,
\end{align}
\end{subequations}
with the nonlinear Poisson constraint $\partial_{xx}\phi=\exp(\phi)-n$ enforced throughout the evolution. This formulation provides a direct way to monitor ion mass and momentum in numerical simulations, while retaining the full nonlinear pressure dependence.

We discretise the fields on a periodic domain of length $L$ using Fourier pseudospectral derivatives and project nonlinear products onto the retained two-thirds Fourier band \cite{Trefethen2000}. We advance $n$ and $m$ using a classical fourth-order Runge-Kutta (RK) scheme \cite{CheneyKincaid2012}. At each RK stage, and after each time step, we obtain the potential by solving the nonlinear Poisson equation with damped Newton iteration and a matrix-free, preconditioned GMRES linear solver \cite{Kelley1995}. We construct initial densities to satisfy the same discretised Poisson constraint, and we choose the initial momentum according to the wave family being tested, for example by imposing the travelling-wave continuity relation or selecting the right-propagating linear eigenmode.

We monitor in numerical simulations the minimum ion density, the Poisson residual, integral mass and momentum drifts, and the resolution of the evolving wave profiles. We use spatial-resolution and time-step tests to separate numerical errors from finite-amplitude departures from the reduced equations.

\section{Linear acoustic response and the origin of nonlinear criticality}
\label{sec:linear}

\subsection{Ion-acoustic dispersion and the long-wave limit}
\label{sec:linear_dispersion}

Before analysing the nonlinear wave regimes, we examine how the prescribed ion EOS modifies the small-amplitude response of the full fP system. This provides both the acoustic reference speed for soliton propagation and the dispersive scale entering the long-wave equations. We introduce perturbations about the homogeneous equilibrium, $n=1+\widetilde n$, $u=\widetilde u$ and $\phi=\widetilde\phi$, with $|\widetilde n|,|\widetilde u|,|\widetilde\phi|\ll1$. Since the electron density satisfies $\exp\phi=1+\widetilde\phi+\mathcal{O}(\widetilde\phi^2)$, the linearised equations become
\begin{subequations}\label{eq:linearised}
\begin{align}
\partial_t\widetilde n+\partial_x\widetilde u&=0,\\
\partial_t\widetilde u+\partial_x\widetilde\phi+\sigma p_1\partial_x\widetilde n&=0,\\
\partial_{xx}\widetilde\phi&=\widetilde\phi-\widetilde n.
\end{align}
\end{subequations}
Notice that only $p_1=p'(1)$ enters this system. The curvature and higher derivatives of the pressure law do not affect the linear waves about the reference state.

To determine the dispersion relation, we seek right-propagating Fourier modes of the form $(\widetilde n,\widetilde u,\widetilde\phi)=(\widehat n,\widehat u,\widehat\phi)\exp[{\mathrm i(kx-\omega t)}]$. The linearised Poisson equation gives $\widehat n=(1+k^2)\widehat\phi$, while continuity gives $\widehat u=(\omega/k)\widehat n$ for $k\ne0$. Substitution into the momentum equation yields
\begin{equation}
\label{eq:dispersion_exact}
\omega^2(k)=k^2\left(\sigma p_1+\frac{1}{1+k^2}\right).
\end{equation}
The two terms inside the parentheses represent the linear ion-pressure response and the electrostatic restoring force, respectively. The latter weakens at shorter wavelengths because electrons screen charge separation over the Debye length. For the right-going branch, we take $\omega_+(k)=k\sqrt{\sigma p_1+(1+k^2)^{-1}}$, with positive frequency for $k>0$.

\begin{figure}[t]
\centering
\includegraphics[width=\linewidth]{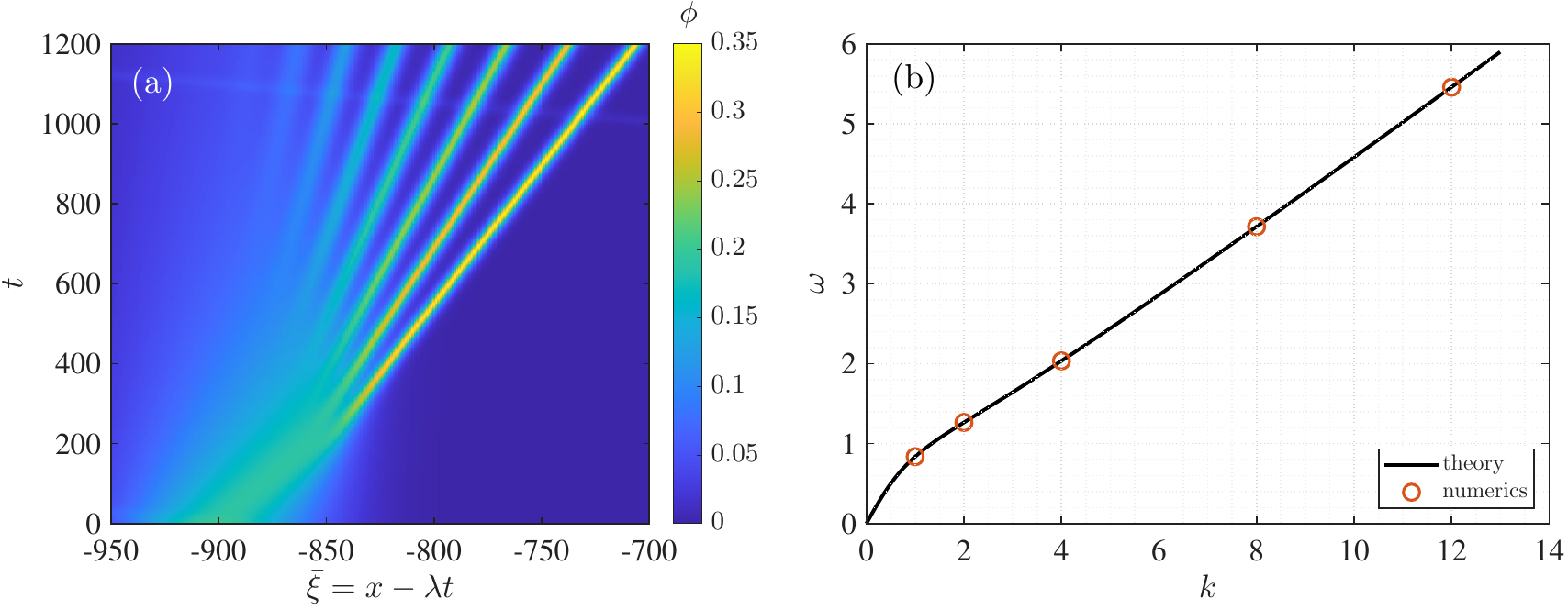}
\caption{\label{fig:linear_and_fission}Nonlinear pulse fission and linear acoustic dispersion in the full fluid-Poisson system. (a) Spatiotemporal evolution of the electrostatic potential in the acoustic frame, $\bar\xi=x-\lambda t$, following a broad Gaussian initial perturbation, $\phi(x,0)=0.2\exp[-((x+900)/45)^2]$. The pulse spontaneously separates into a train of localised structures, with larger-amplitude waves propagating faster relative to the acoustic frame. (b) Exact linear ion-acoustic dispersion relation, Eq.~\eqref{eq:dispersion_exact} (solid curve), compared with numerical frequencies (open circles) obtained from five independent small-amplitude Fourier-mode simulations at $k=1,2,4,8,12$ and fixed first-order density amplitude $A_n=10^{-4}$. Both calculations use $\sigma=0.2$, $p_0=0$, $p_1=1$, $p_2=0.15$ and $p_3=0.02$.}
\end{figure}

The full fP model exhibits dispersive linear propagation and spontaneously generates nonlinear coherent structures under local perturbations. Figure~\ref{fig:linear_and_fission}(a) illustrates the fission of a broad Gaussian potential pulse into a train of localised waves in the acoustic frame. The different trajectory slopes reveal amplitude-dependent propagation speeds. We verified Eq.~\eqref{eq:dispersion_exact} in numerical simulations by exciting individual right-propagating Fourier modes in the nonlinear fP solver. To compare modes at different wavenumbers on the same linear scale, we fixed the first-order ion-density amplitude at $A_n=10^{-4}$ and used the linearised Poisson relation to select the potential amplitude,
\begin{equation}
\label{eq:mode_amplitude}
A_\phi(k)=\frac{A_n}{1+k^2}.
\end{equation}
We initialised the ion momentum using the corresponding right-going eigenvector, suppressing the counter-propagating linear component. After evolving each mode with the full numerical model, we extracted its frequency by linearly fitting the unwrapped temporal phase of its Fourier coefficient.

Figure~\ref{fig:linear_and_fission}(b) compares numerical measurements with Eq.~\eqref{eq:dispersion_exact} for $\sigma=0.2$, $p_1=1$, $p_2=0.15$ and $p_3=0.02$. The numerical frequencies agree with the exact linear prediction to a relative error below $2.5\times10^{-8}$ over the tested modes $k=1,2,4,8,12$. The small-$k$ expansion independently identifies the dispersive coefficient relevant to the weakly nonlinear theory.

The corresponding phase velocity is $v_{\mathrm{ph}}(k)=\omega_+(k)/k$. Thus, in the long-wavelength limit, the acoustic speed is
\begin{equation}
\label{eq:lambda_linear}
\lambda=\lim_{k\to0}v_{\mathrm{ph}}(k)=\sqrt{1+\sigma p_1}.
\end{equation}
A positive $p_1$ therefore increases the ion-acoustic speed relative to the cold-ion value, whereas a negative $p_1$ decreases it, provided $1+\sigma p_1>0$. Expanding Eq.~\eqref{eq:dispersion_exact} for $|k|\ll1$ gives
\begin{equation}
\label{eq:dispersion_longwave}
\omega_+(k)=\lambda k-\frac{k^3}{2\lambda}
+\left(\frac{1}{2\lambda}-\frac{1}{8\lambda^3}\right)k^5+\mathcal{O}(k^7).
\end{equation}
In particular, the leading correction to nondispersive acoustic propagation is $-\gamma k^3$, where
\begin{equation}
\label{eq:gamma_linear}
\gamma=\frac{1}{2\lambda}.
\end{equation}
This coefficient is positive when the long-wave acoustic speed is real and positive. In a frame moving at speed $\lambda$, the frequency becomes $\omega_+(k)-\lambda k=-\gamma k^3+\mathcal{O}(k^5)$, precisely the linear dispersion of the KdV equation \cite{Peyrard2004}. Therefore, the dispersive part of the reduced model follows directly from the screened electrostatic response of the full system, rather than from an independently imposed regularisation.

The condition $\lambda^2>0$ ensures a real acoustic speed as $k\to0$, but it does not, by itself, guarantee stability at every wavelength. Indeed, Eq.~\eqref{eq:dispersion_exact} shows that this local model is linearly stable for all real $k$ if $p_1\geq0$ (with $\sigma>0$). If $p_1<0$, sufficiently large $|k|$ give $\omega^2<0$, even when $\lambda^2>0$. This distinction will be relevant when evaluating strongly correlated pressure closures. For the linear-dispersion and exact-branch validations discussed here, we use $p_1=1$, for which the complete linear branch is stable.

\subsection{Speed of acoustic waves and the fundamental derivative}
\label{sec:linear_fundamental}

The linear dispersion relation identifies the sound speed at one equilibrium state. To understand how weakly nonlinear waves steepen, however, we must also determine how this speed changes when the local density departs from equilibrium. In the quasineutral long-wave limit, Poisson's equation gives $\phi\simeq\ln n$. Consequently, the ion momentum equation reduces to that of a barotropic fluid with the effective pressure
\begin{equation}
\label{eq:effective_pressure}
P_{\mathrm{eff}}(n)=n+\sigma p(n),\qquad
c(n)^2=P_{\mathrm{eff}}'(n)=1+\sigma p'(n).
\end{equation}
The term $n$ is the quasineutral electron contribution, whereas $\sigma p(n)$ retains the prescribed ion response. In particular, $c(1)=\lambda$, so the characteristic speed of the dispersionless model agrees with the long-wave limit of the full fP spectrum.

For the right-going acoustic family, the characteristic velocity is $u+c(n)$. Along a right-going simple wave, $\mbox{d}u/\mbox{d}n=c(n)/n$, and hence its rate of change with compression is
\begin{equation}
\label{eq:characteristic_nonlinearity}
\frac{\mbox{d}}{\mbox{d}n}\bigl[u(n)+c(n)\bigr]
=\frac{c(n)}{n}\,\mathcal G(n),\qquad
\mathcal G(n):=1+\frac{n}{c(n)}\frac{\mbox{d}c(n)}{\mbox{d}n}.
\end{equation}
The quantity $\mathcal G$ is the fundamental derivative of the effective barotropic fluid \cite{Thompson1971Fundamental,CramerKluwick1984}. Its sign determines whether the right-going characteristic speed increases or decreases along a compressive simple wave. At $\mathcal G=0$, the first-order variation vanishes, corresponding to the local loss of genuine nonlinearity of this characteristic field \cite{Lax1957Hyperbolic}.

Using Eq.~\eqref{eq:effective_pressure}, the fundamental derivative can be expressed directly in terms of the ion pressure law,
\begin{equation}
\label{eq:fundamental_derivative}
\mathcal G(n)=1+\frac{\sigma n p''(n)}{2[1+\sigma p'(n)]}.
\end{equation}
At the reference state, the Taylor convention of Eq.~\eqref{eq:pressure_expansion} gives
\begin{equation}\label{eq:fundamental_at_equilibrium}
\mathcal G(1)=\frac{1+\sigma(p_1+p_2)}{\lambda^2}.
\end{equation}
This result makes the different roles of pressure slope and curvature explicit. Parameter $p_1$ fixes the acoustic speed, whereas $p_2$ controls the first density variation of the characteristic speed. In the weakly nonlinear reduction developed in the next section, the quadratic KdV coefficient is $\alpha=\lambda\mathcal G(1)$. Thus, its vanishing is equivalent to $\mathcal G(1)=0$, identifying a critical line where the dispersionless acoustic field loses its genuine nonlinearity. The pressure's higher derivatives then determine the leading nonlinear behaviour at this degeneracy.

\section{EOS-induced KdV-Gardner criticality}
\label{sec:nonlinear}

We now examine how changes in the ion-pressure curvature modify the nonlinear acoustic response while preserving the underlying long-wave branch. The linear-wave analysis identifies the acoustic branch speed and the dispersive correction produced by charge separation. It also shows that the density dependence of the acoustic speed is governed by the fundamental derivative $\mathcal G(n)$, defined in Eq.~\eqref{eq:fundamental_derivative}. We now examine how these two ingredients enter the weakly nonlinear dynamics. Next, we derive the KdV, mKdV, and Gardner equations as successive regimes of the same acoustic characteristic field as its quadratic nonlinearity approaches zero.

\subsection{Quadratic acoustic nonlinearity and the KdV regime}\label{sec:nonlinear_kdv}

For a weak disturbance propagating to the right, the long-wavelength limit is characterised by a small potential amplitude and a slow evolution relative to the acoustic frame. We introduce a small and positive dimensionless parameter $\varepsilon\ll1$ and the following stretched coordinates,
\begin{equation}
\label{eq:kdv_scaling}
\xi:=\varepsilon^{1/2}(x-\lambda t),\qquad
\tau:=\varepsilon^{3/2}t,
\end{equation}
where $\lambda=c(1)$ is the acoustic speed given by Eq.~\eqref{eq:lambda_linear}. We expand the fields in powers of $\varepsilon$ as follows, 
\begin{equation}
\label{eq:kdv_expansion}
\begin{aligned}
n&=1+\varepsilon n_1+\varepsilon^2 n_2+\mathcal{O}(\varepsilon^3),\\
u&=\varepsilon u_1+\varepsilon^2u_2+\mathcal{O}(\varepsilon^3),\\
\phi&=\varepsilon\phi_1+\varepsilon^2\phi_2+\mathcal{O}(\varepsilon^3).
\end{aligned}
\end{equation}
At leading order, Poisson's equation imposes $n_1=\phi_1$, while continuity and momentum give $u_1=\lambda\phi_1$ and $\lambda^2=1+\sigma p_1$. The three perturbations therefore belong to the same right-propagating acoustic mode. At the next non-trivial order, elimination of $n_2$, $u_2$ and $\phi_2$ leads to the solvability condition
\begin{equation}
\label{eq:kdv_solvability}
2\lambda\,\partial_\tau\phi_1+
2\bigl[1+\sigma(p_1+p_2)\bigr]\phi_1\partial_\xi\phi_1
+\partial_{\xi\xi\xi}\phi_1=0.
\end{equation}
Thus, the leading potential amplitude satisfies the KdV equation \cite{Taniuti1968, Taniuti1974},
\begin{equation}
\label{eq:kdv_reduced}
\partial_\tau\phi_1+\alpha\phi_1\partial_\xi\phi_1
+\gamma\partial_{\xi\xi\xi}\phi_1=0,\qquad
\alpha:=\frac{1+\sigma(p_1+p_2)}{\lambda}.
\end{equation}
The dispersive coefficient $\gamma$ agrees with Eq.~\eqref{eq:gamma_linear}, following the small-wavenumber expansion in Eq.~\eqref{eq:dispersion_longwave}. Importantly, the nonlinear coefficient admits a direct interpretation through the fundamental derivative. From Eq.~\eqref{eq:fundamental_at_equilibrium},
\begin{equation}
\label{eq:alpha_fundamental}
\alpha=\lambda\mathcal G(1).
\end{equation}
Equation~\eqref{eq:alpha_fundamental} specifies, for the present electrostatic closure, the general connection between weakly nonlinear steepening and genuine nonlinearity that also arises in dispersive Eulerian fluids \cite{Hoefer2014DispersiveEulerian}. In our normalisation, $\alpha$ equals the first potential-amplitude derivative of the right-going acoustic characteristic speed at equilibrium. For $\alpha>0$, the familiar small-amplitude solitons are compressive. For $\alpha<0$, they are rarefactive. As $\alpha$ approaches zero, however, the quadratic balance used in Eq.~\eqref{eq:kdv_scaling} becomes non-uniform, and the higher-order nonlinearity must be retained.

\subsection{Loss of genuine nonlinearity and the cubic critical point}
\label{sec:nonlinear_mkdv}

Because $\lambda>0$, the quadratic coefficient vanishes if and only if the equilibrium fundamental derivative vanishes. Using the pressure expansion of Eq.~\eqref{eq:pressure_expansion}, the critical manifold is
\begin{equation}
\label{eq:critical_manifold}
\alpha=0\quad\Longleftrightarrow\quad
\mathcal G(1)=0\quad\Longleftrightarrow\quad
1+\sigma(p_1+p_2)=0,\qquad
p_{2,c}:=-p_1-\frac{1}{\sigma}.
\end{equation}
At this degeneracy, the acoustic field ceases to be genuinely nonlinear at the reference state. Its characteristic speed no longer changes linearly with the wave amplitude, so the next-order density derivative of $\mathcal G$ controls the leading steepening. This is also why taking $\alpha\to0$ in the finite-amplitude KdV soliton formula does not provide a consistent critical pulse.

The cubic mechanism follows directly from the simple-wave characteristics introduced in \S\ref{sec:linear_fundamental}. In the quasineutral approximation $n=\exp\phi$, a right-going simple wave obeys $\partial_t\phi+\Lambda_+(\phi)\partial_x\phi=0$, where $\Lambda_+(\phi)=u(\exp\phi)+c(\exp\phi)$ and $\mbox{d}u/\mbox{d}n=c(n)/n$. It follows that
\begin{equation}
\label{eq:char_phi_derivatives}
\frac{\mbox{d}\Lambda_+}{\mbox{d}\phi}=c(n)\mathcal G(n),\qquad
\left.\frac{\mbox{d}^2\Lambda_+}{\mbox{d}\phi^2}\right|_{\phi=0}
=\lambda\mathcal G'(1)\quad\text{when }\mathcal G(1)=0.
\end{equation}
Hence, at the critical point, the characteristic speed has the expansion
\begin{equation}
\label{eq:char_critical_expansion}
\Lambda_+(\phi)=\lambda+\frac{\lambda}{2}\mathcal G'(1)\phi^2
+\mathcal{O}(\phi^3).
\end{equation}
The first nonlinear correction is now quadratic in $\phi$, rather than linear. Combining this cubic steepening with the dispersive correction requires the longer spatial and temporal scales
\begin{equation}
\label{eq:mkdv_scaling}
\xi=\varepsilon(x-\lambda t),\qquad
\tau=\varepsilon^3t,\qquad \phi=\varepsilon\phi_1+\mathcal{O}(\varepsilon^2).
\end{equation}
The same fP perturbation hierarchy then gives, at the first non-vanishing evolution order, the mKdV equation
\begin{equation}
\label{eq:mkdv_reduced}
\partial_{\tau}\phi_1
+\beta_c\phi_1^2\partial_{\xi}\phi_1
+\gamma\partial_{\xi\xi\xi}\phi_1=0,\qquad
\beta_c:=\frac{\lambda}{2}\mathcal G'(1).
\end{equation}
The subscript in $\beta_c$ emphasises that this coefficient is evaluated on the critical manifold. From Eq.~\eqref{eq:fundamental_derivative},
\begin{equation}
\label{eq:Gprime_equilibrium}
\mathcal G'(1)=\frac{\sigma(p_2+3p_3)}{\lambda^2}
-\frac{2\sigma^2p_2^2}{\lambda^4}.
\end{equation}
Using $\sigma p_{2,c}=-\lambda^2$, we obtain
\begin{equation}
\label{eq:beta_critical}
\beta_c=-\lambda+\frac{\sigma}{2\lambda}(p_{2,c}+3p_3)
=\frac{3}{2\lambda}\bigl(\sigma p_3-\lambda^2\bigr).
\end{equation}

Thus, the sign of the cubic term is determined by how the fundamental derivative crosses zero. For $\gamma>0$, $\beta_c>0$ gives a focusing mKdV reduction that supports localised, zero-background solitons of either polarity. On the other hand, $\beta_c<0$ produces the defocusing reduction, for which the corresponding zero-background bright pulses do not exist, although non-zero-background kink-type structures may be admissible. If both $\mathcal G(1)=0$ and $\mathcal G'(1)=0$, the cubic term also vanishes. In the present Taylor parametrisation, this higher degeneracy occurs at $p_2=-p_1-1/\sigma$ and $p_3=p_1+1/\sigma$, and lies beyond the mKdV approximation. The analogous simultaneous cancellation of the quadratic and cubic terms has been studied at supercritical plasma compositions \cite{VerheestOlivierHereman2016}. The condition above identifies its EOS-controlled counterpart, whose leading reduced evolution is derived in \S\,\ref{sec:nonlinear_hierarchy}.

\subsection{A two-sided unfolding: the Gardner regime}
\label{sec:nonlinear_gardner}

The criticality condition fixes the pressure curvature at $p_{2,c}$. To describe nearby states, we introduce an unfolding of this degeneracy. In bifurcation theory, an unfolding embeds a degenerate problem in a parameter-dependent family, revealing how distinct behaviours emerge as the control parameter moves away from its critical value \cite{GolubitskySchaeffer1985}. Here, quadratic and cubic nonlinearities must remain comparable, so we perturb the pressure curvature on the same scale as the wave amplitude. We therefore write
\begin{equation}
\label{eq:gardner_detuning}
p_2=p_{2,c}+\varepsilon q,\qquad
\alpha=\varepsilon a,\qquad
a:=\frac{\sigma q}{\lambda},
\end{equation}
with $q=\mathcal{O}(1)$ of either sign. Keeping the critical scales of Eq.~\eqref{eq:mkdv_scaling} and expanding consistently through the evolution order yields
\begin{equation}
\label{eq:gardner_scaled}
\partial_{\tau}\phi_1+a\phi_1\partial_{\xi}\phi_1
+\beta_c\phi_1^2\partial_{\xi}\phi_1
+\gamma\partial_{\xi\xi\xi}\phi_1=0,
\end{equation}
which is the Gardner equation \cite{VerheestHereman2025}. We remark that the coefficient appearing in this scaled equation is $a=\alpha/\varepsilon$, not the small physical coefficient $\alpha$ itself. This distinction is necessary because $\alpha=\mathcal{O}(\varepsilon)$ in the near-critical ordering. Equation~\eqref{eq:gardner_scaled} applies for both $q>0$ and $q<0$, with mKdV recovered continuously at $q=0$. Its nonlinear flux,
\begin{equation}\label{eq:gardner_flux}
\mathcal F(\phi_1)=\frac{a}{2}\phi_1^2+\frac{\beta_c}{3}\phi_1^3,
\end{equation}
provides a local dispersive regularisation of the non-convex acoustic characteristic law when the quadratic contribution is sufficiently weak.

The relative importance of the cubic term can be expressed directly in terms of the peak physical potential amplitude $A=\mathcal{O}(\varepsilon)$. At leading order, the two contributions to the soliton speed excess scale as $\alpha A/3$ and $\beta_c A^2/6$. Their ratio defines a crossover measure,
\begin{equation}
\label{eq:crossover_parameter}
\mathcal C:=\left|\frac{\beta_c A}{2\alpha}\right|,\qquad \alpha\ne0.
\end{equation}
The quadratic KdV term dominates when $\mathcal C\ll1$, whereas the cubic contribution becomes equally important when $\mathcal C=O(1)$. Near $\alpha=0$, the Gardner description is therefore required on either side of the critical manifold. The sign of $\beta_c$ remains a separate property of the selected pressure law and controls which travelling-wave branches can occur.

\subsection{Soliton scaling and the transition between regimes}
\label{sec:nonlinear_solitary}

The different asymptotic balances lead to distinct relations between wave amplitude, velocity and width. For a travelling profile of the form $\phi_1(\xi,\tau)=F(z_s)$ with $z_s:=\xi-U_s\tau$ and zero far-field background, integration of Eq.~\eqref{eq:gardner_scaled} gives
\begin{equation}
\label{eq:gardner_travelling_ode}
\gamma \frac{\mbox{d}^2F}{\mbox{d}z_s^2}=U_sF-\frac{a}{2}F^2-\frac{\beta_c}{3}F^3.
\end{equation}
A second integration, with $F,\mbox{d}F/\mbox{d}z_s\to0$ at infinity, gives
\begin{equation}
\label{eq:gardner_energy}
\frac{\gamma}{2}\left(\frac{\mbox{d}F}{\mbox{d}z_s}\right)^2=
\frac{U_s}{2}F^2-\frac{a}{6}F^3-\frac{\beta_c}{12}F^4.
\end{equation}
The turning-point condition $F(0)=A_s$ and $\left.\mbox{d}F/\mbox{d}z_s\right|_{z_s=0}=0$ then yields
\begin{equation}
\label{eq:gardner_speed_scaled}
U_s=\frac{a}{3} A_s+\frac{\beta_c}{6}A_s^2.
\end{equation}
Because the physical wave amplitude is $A=\varepsilon A_s$ and the physical velocity excess over $\lambda$ is $V=\varepsilon^2U_s$, this simply becomes
\begin{equation}
\label{eq:gardner_speed_physical}
V=\frac{\alpha}{3}A+\frac{\beta_c}{6}A^2,
\end{equation}
to leading near-critical order. The first term is the familiar KdV amplitude-speed relation, while the second survives at the critical point and is even in the signed amplitude. It therefore predicts equal leading-order speed excesses for positive and negative mKdV pulses at equal values of $|A|$, leaving finite-amplitude polarity asymmetries to higher-order corrections. Opposite-polarity near-critical pulses have previously been described through Sagdeev expansions for composition-tuned plasmas \cite{OlivierVerheestMaharaj2016}. Our amplitude-speed relation displays the corresponding response when the control parameter is the ion-pressure curvature.

The zero-background KdV soliton has the form $\phi\simeq A\operatorname{sech}^2[(x-(\lambda+V)t)/\Delta]$, with $V=\alpha A/3$ and $\Delta=(4\gamma/V)^{1/2}$, provided $V>0$. At criticality, the focusing mKdV pulse is $\phi\simeq A\operatorname{sech}[(x-(\lambda+V)t)/\Delta]$, with $V=\beta_c A^2/6$ and $\Delta=(\gamma/V)^{1/2}$, requiring $\beta_c>0$. These relations distinguish the KdV width law $\Delta\propto|A|^{-1/2}$ from the critical mKdV law $\Delta\propto|A|^{-1}$.

For the near-critical Gardner equation, the corresponding localised profile can be written in the physical variables as follows:
\begin{equation}
\label{eq:gardner_profile_physical}
\phi(x,t)\simeq\frac{A(2\alpha+\beta_c A)}{\alpha+(\alpha+\beta_c A)
\cosh\!\left[\sqrt{V/\gamma}\,(x-(\lambda+V)t-x_0)\right]},
\end{equation}
where $V$ is given by Eq.~\eqref{eq:gardner_speed_physical}. This expression requires $V>0$ and a nonsingular denominator. The condition $(\alpha+\beta_c A)(2\alpha+\beta_c A)>0$ ensures a regular exponentially localised branch away from limiting cases. It reduces to the KdV profile when the cubic contribution is negligible, and to the focusing mKdV profile as $\alpha\to0$ with $\beta_c>0$. In the defocusing case, the admissible branches and their possible non-zero backgrounds require a separate travelling-wave analysis, rather than continuation of the bright-pulse formula through $\beta_c=0$.

\subsection{Higher-order constitutive degeneracy and the evolution hierarchy}
\label{sec:nonlinear_hierarchy}

The critical reductions admit a systematic extension on exactly tuned constitutive manifolds. For a fixed integer $k\geq0$, suppose that
$\mathcal G^{(j)}(1)=0$ for $0\leq j<k$, while $\mathcal G^{(k)}(1)\neq0$. A unidirectional long-wave expansion of the full fP equations with $\phi=\varepsilon\phi_1+\cdots$ requires $\xi=\varepsilon^{(k+1)/2}(x-\lambda t)$ and
$\tau=\varepsilon^{3(k+1)/2}t$. The leading solvability condition is, therefore,
\begin{equation}\label{eq:evolution_hierarchy}
 \partial_\tau\phi_1
 +\frac{\lambda\mathcal G^{(k)}(1)}{(k+1)!}
 \phi_1^{k+1}\partial_\xi\phi_1
 +\gamma\partial_{\xi\xi\xi}\phi_1=0,
 \qquad \gamma=\frac{1}{2\lambda}.
\end{equation}
Equation~\eqref{eq:evolution_hierarchy} reduces to KdV for $k=0$ and mKdV for $k=1$. In the pressure Taylor expansion, the successive cancellations require $p_j=(-1)^{j-1}\lambda^2/\sigma$ for
$2\leq j\leq k+1$, with the set empty for $k=0$. We emphasise that Eq.~\eqref{eq:evolution_hierarchy} applies at an exact degeneracy, and its detuned counterpart generally contains mixed nonlinearities, as in the Gardner regime. Interestingly, the entire criticality hierarchy admits a closed-form constitutive limit. The condition $\mathcal G(n)= 0$ implies $n c(n)=\lambda$, which yields the effective Chaplygin-von K\'arm\'an-Tsien pressure law \cite{Serre2011Chaplygin,KongWei2014Chaplygin},
\begin{equation}
P_{\mathrm{eff}}(n)=P_0-\frac{\lambda^2}{n}.
\end{equation}
This classical linearly degenerate acoustic closure reproduces the hierarchy of critical Taylor coefficients $p_j=(-1)^{j-1}\lambda^2/\sigma$ for every $j\geq 2$, together with $p_1=(\lambda^2-1)/\sigma$. The successive cancellations of acoustic nonlinearities can therefore be understood as finite-order realisations of an exact, linearly degenerate constitutive law.

At the codimension-two state, $\mathcal G(1)=\mathcal G'(1)=0$, write $p_4:=p^{(4)}(1)/4!$ and define $\delta_4:=\lambda\mathcal G''(1)/6$. Then
\begin{equation}
\label{eq:quartic_evolution}
 \partial_\tau\phi_1+\delta_4\phi_1^3\partial_\xi\phi_1
 +\gamma\partial_{\xi\xi\xi}\phi_1=0,
 \qquad
 \delta_4=\frac{2(\lambda^2+\sigma p_4)}{\lambda},
\end{equation}
with the critical scales $\xi=\varepsilon^{3/2}(x-\lambda t)$ and $\tau=\varepsilon^{9/2}t$. For $\delta_4 A^3>0$, this reduced equation
has a small-amplitude, zero-background soliton with leading physical speed excess and profile
\begin{equation}\label{eq:quartic_profile}
 V=\frac{\delta_4}{10}A^3
 =\frac{\lambda^2+\sigma p_4}{5\lambda}A^3,
 \qquad
 \Phi(z)=A\operatorname{sech}^{2/3}\!\left(
 \frac32\sqrt{\frac{V}{\gamma}}\,z\right),
 \quad z=x-(\lambda+V)t.
\end{equation}
As we will see later in \S\,\ref{sec:sagdeev_critical}, the speed law of Eq.~\eqref{eq:quartic_profile} agrees with the independent exact Sagdeev expansion. Quartic-nonlinearity reductions at supercritical plasma compositions have been obtained previously \cite{VerheestOlivierHereman2016}. Here, we identify the corresponding coefficient and scaling for EOS-controlled criticality.

The hierarchy shows how the first non-vanishing density derivative of $\mathcal G$ selects the leading nonlinear amplitude equation at each exact constitutive degeneracy. The resulting small-amplitude solitons must nevertheless be compared against the exact stationary equations. In the next section we examine their full fP-Sagdeev representation and the admissibility of the associated spatial branches.

\section{Exact travelling-wave formulation and Sagdeev criticality}
\label{sec:sagdeev}

The reduced KdV-Gardner equations describe solitons when their amplitude and spatial gradients are sufficiently small. To determine which of these structures are supported by the parent fP model, we now consider travelling waves without imposing a weak-amplitude expansion. This provides a description of the nonlinear wave branches and, in particular, allows us to identify the quadratic degeneracy directly from the exact spatial dynamics.

\subsection{Travelling-wave invariants and the Sagdeev pseudopotential}\label{sec:sagdeev_invariants}

We seek stationary profiles in the coordinate $\zeta:=x-Mt$, where $M>0$ is the propagation speed in the laboratory frame. Under this ansatz, $n=n(\zeta)$,  $u=u(\zeta)$ and $\phi=\phi(\zeta)$ for fixed $M$. For a localised profile on the reference background, the boundary conditions are $n\to1$, $u\to0$ and $\phi\to0$, together with vanishing spatial derivatives, as $|\zeta|\to\infty$. The continuity equation~\eqref{eq:fp_continuity} can be integrated once to give
\begin{equation}
\label{eq:sagdeev_continuity}
 n(u-M)=-M,\qquad u(n)=M\left(1-\frac{1}{n}\right).
\end{equation}
Thus, density and velocity are not independent along a travelling wave. The ion flux relative to the wave remains constant. Substituting this relation into the momentum equation and integrating from the reference state, we obtain the exact Bernoulli relation
\begin{equation}
\label{eq:sagdeev_bernoulli}
 \phi(\zeta)=\mathcal B(n(\zeta);M):=\frac{M^2}{2}\left(1-\frac{1}{n^2}\right)
 -\sigma\int_1^n\frac{1}{s}\frac{\mbox{d}p}{\mbox{d}s}\,\mbox{d}s.
\end{equation}
The first term accounts for the change in ion kinetic energy in the travelling frame, whereas the second describes the ion-pressure contribution for the prescribed barotropic closure. In particular, the Bernoulli relation~\eqref{eq:sagdeev_bernoulli} does not require a polynomial pressure law. Its derivative,
\begin{equation}
\label{eq:sagdeev_density_map}
 \partial_n\mathcal B=\frac{M^2-\sigma n^2p'(n)}{n^3},
\end{equation}
shows where the density-potential mapping is locally invertible. We restrict attention to positive densities and regular branches for which $\partial_n\mathcal B\ne0$ along the profile. A zero of this derivative marks a singularity of the density-potential inversion in the stationary fluid reduction and cannot be crossed using the present single-valued construction.

On such a branch, Eq.~\eqref{eq:sagdeev_bernoulli} defines $n=n(\mathcal B;M)$. Poisson's equation becomes a second-order autonomous equation for the electrostatic potential,
\begin{equation}
\label{eq:sagdeev_ode}
 \frac{\mbox{d}^2\phi}{\mbox{d}\zeta^2}=e^\phi-n(\phi;M)
 =-\partial_{\phi}\mathcal S(\phi;M),
\end{equation}
where
\begin{equation}
\label{eq:sagdeev_potential}
\mathcal S(\phi;M)=\int_0^\phi
 \big[n(\psi;M)-e^\psi\big]\,\mbox{d}\psi,
\end{equation}
is the Sagdeev pseudopotential governing the spatial profile \cite{Sagdeev1966}. The Chaplygin-von K\'arm\'an-Tsien closure identified
in \S~\ref{sec:nonlinear_hierarchy} provides an exact limiting case of this stationary formulation. At the acoustic speed $M=\lambda$, the Bernoulli relation reduces to $\mathcal B(n;\lambda)=\ln n$, and the corresponding Sagdeev pseudopotential vanishes identically, $\mathcal S(\phi;\lambda)= 0$. Thus, the complete cancellation of acoustic nonlinearities in the evolution hierarchy has an exact counterpart in the stationary fP system.

Multiplying Eq.~\eqref{eq:sagdeev_ode} by $\mbox{d}\phi/\mbox{d}\zeta$ and using the far-field conditions yields the spatial energy integral
\begin{equation}
\label{eq:sagdeev_energy}
\frac12\left(\frac{\mbox{d}\phi}{\mbox{d}\zeta}\right)^2+\mathcal S(\phi;M)=0.
\end{equation}
Within this formulation, the pseudopotential can be computed parametrically without explicitly inverting $\mathcal B(n;M)$. Indeed, inserting Eq.~\eqref{eq:sagdeev_density_map} into Eq.~\eqref{eq:sagdeev_potential} and integrating gives
\begin{equation}
\label{eq:sagdeev_parametric}
 \mathcal S\big(\mathcal B(n;M);M\big)
 =M^2\left(1-\frac1n\right)
 -\sigma\big[p(n)-p(1)\big]
 -\big(e^{\mathcal B(n;M)}-1\big).
\end{equation}
Equations~\eqref{eq:sagdeev_bernoulli} and \eqref{eq:sagdeev_parametric} provide an exact travelling-wave representation for an arbitrary sufficiently smooth $p(n)$, which we compare later in this article with numerical profiles obtained from the full time-dependent equations.

\subsection{Spatial equilibria and the critical pseudopotential}
\label{sec:sagdeev_critical}

The far-field state is an equilibrium of the spatial dynamics because $\mathcal S(0;M)=\partial_\phi\mathcal S(0;M)=0$. A zero-background soliton with a signed turning-point potential $A\ne0$ requires a second, simple root $\mathcal S(A;M)=0$, with $\partial_\phi\mathcal S(A;M)\ne0$, and $\mathcal S(\phi;M)<0$ between $0$ and $A$. The resulting profile is a homoclinic orbit in the phase plane $(\phi,\mbox{d}\phi/\mbox{d}\zeta)$. This condition distinguishes an admissible soliton from an algebraic root of the amplitude-speed relation alone. The pseudopotential must remain accessible along the entire orbit, and the density map must remain regular.

The nature of the reference equilibrium follows from the local curvature of $\mathcal S$. Using Eq.~\eqref{eq:sagdeev_density_map} at $n=1$, we obtain
\begin{equation}
\label{eq:sagdeev_second}
 \partial_{\phi\phi}\mathcal S(0;M)
 =\frac{1}{M^2-\sigma p_1}-1.
\end{equation}
For the regular acoustic branch connected continuously to $M=\lambda$, the denominator is positive. The equilibrium is a saddle when $\partial_{\phi\phi}\mathcal S(0;M)<0$, which requires $M>\lambda$ for right-moving waves. At $M=\lambda$, the curvature vanishes, signalling the small-amplitude limit in which nonlinear corrections determine the spatial orbit.

Higher derivatives of the exact pseudopotential reveal the same criticality as the fundamental derivative. Differentiating Eq.~\eqref{eq:sagdeev_density_map} twice, and evaluating at the acoustic speed $M=\lambda$, gives
\begin{equation}
\label{eq:sagdeev_third}
 \partial_{\phi\phi\phi}\mathcal S(0;\lambda)
 =2\big[1+\sigma(p_1+p_2)\big]
 =2\lambda\alpha
 =2\lambda^2\mathcal G(1).
\end{equation}
Thus, the criticality condition $\mathcal G(1)=0$ has a direct signature in the exact fP travelling-wave problem. The cubic term in the Sagdeev pseudopotential vanishes at the acoustic threshold. Near-critical Sagdeev–Taylor expansions, and how they tie into reduced evolution equations, are well known in plasma theory \cite{OlivierVerheestMaharaj2016, OlivierVerheestMaharaj2016Corrigendum, OlivierVerheestMaharaj2017}. In this work, we establish the explicit equivalence between this coefficient and the fundamental derivative of the effective pressure for a general, smooth local ion closure, assuming the density–potential relation is regular.

At the critical point, the next derivative is
\begin{equation}\label{eq:sagdeev_fourth}
 \left.\partial_{\phi\phi\phi\phi}\mathcal S(0;\lambda)\right|_{\alpha=0}
 =6\big(\sigma p_3-\lambda^2\big)
 =4\lambda\beta_c
 =2\lambda^2\mathcal G'(1).
\end{equation}
Consequently, the cubic mKdV coefficient can also be read directly from the quartic term of the exact pseudopotential. The three descriptions (acoustic characteristics, reductive perturbation theory and Sagdeev spatial dynamics) therefore identify the same critical manifold and the same sign of the leading critical nonlinearity.

The connection between the acoustic fundamental derivative and the exact Sagdeev potential extends beyond the first two degeneracies. Suppose the EOS is sufficiently smooth and, for some non-negative integer $k$, $\mathcal G(1)=\mathcal G'(1)=\cdots=\mathcal G^{(k-1)}(1)=0$ (for $k=0$, the conditions are trivially satisfied). Then the first potentially nonzero higher derivative obeys the exact identity
\begin{equation}\label{eq:sagdeev_hierarchy}
\left.\frac{\partial^{k+3}}{\partial\phi^{k+3}}\mathcal S(\phi;M)\right|_{\phi=0,\,M=\lambda}
=2\lambda^2\mathcal G^{(k)}(1).
\end{equation}
The cases $k=0$ and $k=1$ recover Eqs.~\eqref{eq:sagdeev_third} and \eqref{eq:sagdeev_fourth}, respectively.

At the simultaneous quadratic and cubic degeneracy, write $p_4:=p^{(4)}(1)/4!$. For the quartic pressure expansion with $p_2=-\lambda^2/\sigma$ and $p_3=\lambda^2/\sigma$, one obtains
\begin{equation}\label{eq:sagdeev_fifth}
\partial_{\phi\phi\phi\phi\phi}\mathcal S(0;\lambda)
=2\lambda^2\mathcal G''(1)
=24(\lambda^2+\sigma p_4).
\end{equation}
Accordingly, if this coefficient is nonzero, the exact small-amplitude zero-background turning-point condition gives $V=(\lambda^2+\sigma p_4)A^3/(5\lambda)+o(A^3)$ as $A\to0$ and $V=M-\lambda\to0^+$. At this order, a sufficiently small pulse can therefore have only the polarity selected by the sign of $\lambda^2+\sigma p_4$. This stationary coefficient agrees with the independently derived quartic-nonlinearity evolution equation \eqref{eq:quartic_evolution} and the leading profile \eqref{eq:quartic_profile}.

\subsection{Recovery of the Gardner normal form and soliton branches}
\label{sec:sagdeev_gardner}

To connect the exact description with the weakly nonlinear theory, consider speeds $M=\lambda+V$ and potential amplitudes $\phi=\mathcal{O}(\varepsilon)$, with $V=\mathcal{O}(\varepsilon^2)$ and $\alpha=\mathcal{O}(\varepsilon)$ near the critical manifold. Expanding Eqs.~\eqref{eq:sagdeev_second}-\eqref{eq:sagdeev_fourth} consistently gives
\begin{equation}\label{eq:sagdeev_expansion}
 \mathcal S(\phi;\lambda+V)
 =-\lambda V\phi^2
 +\frac{\lambda\alpha}{3}\phi^3
 +\frac{\lambda\beta_c}{6}\phi^4
 +\mathcal{O}(\varepsilon^5),
\end{equation}
where the remainder is understood under the stated joint scaling. The quadratic contribution describes the speed excess above the acoustic threshold, while the cubic and quartic contributions are controlled by $\mathcal G(1)$ and $\mathcal G'(1)$, respectively. Substituting Eq.~\eqref{eq:sagdeev_expansion} into Eq.~\eqref{eq:sagdeev_ode}, and using $\gamma=(2\lambda)^{-1}$, leads to
\begin{equation}
\label{eq:sagdeev_gardner_ode}
 \gamma\frac{\mbox{d}^2\phi}{\mbox{d}\zeta^2}=V\phi-\frac{\alpha}{2}\phi^2
 -\frac{\beta_c}{3}\phi^3+\cdots,
\end{equation}
which is precisely the travelling-wave equation of the Gardner reduction in physical variables. Thus, the Gardner normal form is the local expansion of the exact Sagdeev spatial dynamics.

At a non-zero pulse maximum or minimum, $\phi=A$ and $\mbox{d}\phi/\mbox{d}\zeta=0$. The first non-trivial turning-point condition obtained from Eq.~\eqref{eq:sagdeev_expansion} at leading near-critical order recovers Eq.~\eqref{eq:gardner_speed_physical}, which shows how the exact pseudopotential organises the polarity-dependent soliton branches. In particular, at $\alpha=0$ a focusing cubic term ($\beta_c>0$) gives two small-amplitude zero-background turning points with $A\simeq\pm\sqrt{6V/\beta_c}$. A defocusing cubic term ($\beta_c<0$) cannot balance the negative quadratic pseudopotential for $V>0$ and therefore does not support these zero-background bright pulses near the acoustic threshold. Instead, its non-zero-background structures, including possible fronts and double layers, must be considered with their appropriate asymptotic states and spatial energy level. They are not described by the zero-background homoclinic condition alone.

Beyond the small-amplitude limit, the exact amplitude-speed relation is obtained by solving
\begin{equation}
\label{eq:sagdeev_exact_turning}
 \mathcal S\big(\mathcal B(n_A;M);M\big)=0,
 \qquad A=\mathcal B(n_A;M),
\end{equation}
subject to $n_A>0$, the sign condition on $\mathcal S$ along the orbit and the regularity requirement in Eq.~\eqref{eq:sagdeev_density_map}. This formulation is particularly useful for assessing systematic finite-amplitude deviations from the KdV-Gardner velocity predictions in the full numerical simulations. A systematic phase-plane classification can therefore distinguish ordinary homoclinic pulses, limiting double roots and heteroclinic fronts, without identifying spatial saddle or centre equilibria as temporally stable or unstable plasma waves.

To verify finite-time propagation of selected stationary solutions, we initialised the fP equations using profiles constructed from the exact Sagdeev pseudopotential. At $t=120$, these structures exhibit negligible measured amplitude changes, with speed discrepancies relative to the exact stationary predictions no larger than $3.22\times10^{-8}$. In contrast, initial conditions obtained from the asymptotic mKdV and Gardner profiles deform and may exhibit speed discrepancies substantial relative to the low excess speed $M-\lambda$. Interestingly, these discrepancies depend on the procedure used to identify the pulse centre, indicating that the propagation of an evolving approximate profile should not be confused with the velocity of an exact soliton.

\section{EOS-controlled soliton bifurcations}
\label{sec:bifurcations}

The preceding sections identified the loss of genuine acoustic nonlinearity as a degeneracy of the local wave equation and of the exact Sagdeev pseudopotential. We now ask how this degeneracy organises the families of travelling waves that are admissible in the parent fP system. The EOS provides two independent controls: its curvature $p_2$ determines the departure from quadratic criticality, whereas $p_3$ fixes the sign of the cubic nonlinearity on the critical manifold. To separate these effects, we first vary one coefficient while keeping the other fixed. Throughout this section, we use the polynomial pressure closure in Eq.~\eqref{eq:pressure_expansion} as a controlled constitutive example.

We choose $\sigma=0.2$ and $p_1=1$, so that $\lambda=\sqrt{1.2}$ and $p_{2,c}=-6$. The linear spectrum of the prescribed local fP model is stable for this reference pressure slope. For every proposed zero-background pulse, we obtain its signed peak potential $A$ from the exact turning-point condition \eqref{eq:sagdeev_exact_turning}. We then check that the pseudopotential remains negative between the equilibrium and the turning point, and that the density-potential mapping is regular along the orbit. Only roots satisfying these conditions are retained in the bifurcation diagrams. The exact curves therefore represent admissible homoclinic orbits.

\subsection{Unfolding the quadratic criticality through the pressure curvature}\label{sec:bifurcations_curvature}

\begin{figure}[tbp]
 \centering
 \includegraphics[width=\linewidth]{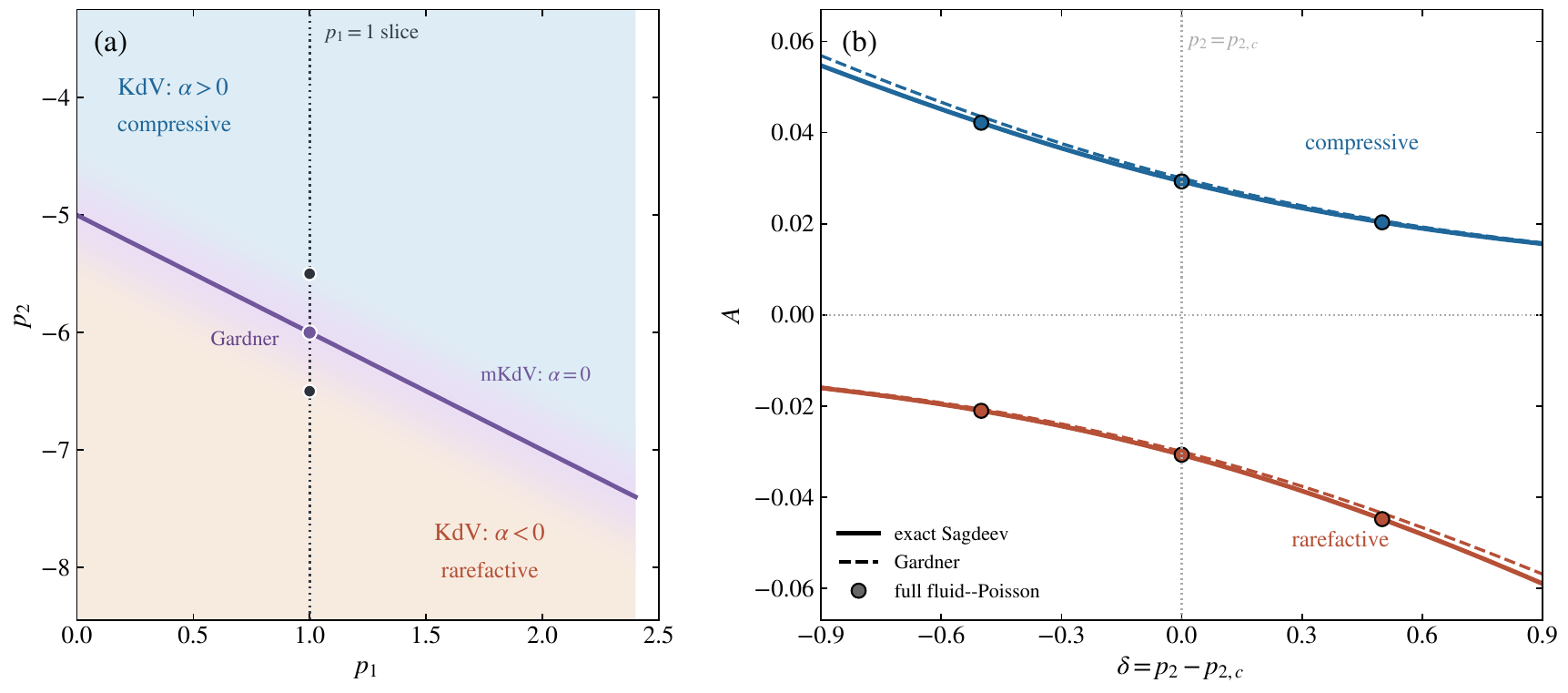}
 \caption{Equation-of-state control of quadratic acoustic criticality and soliton branches. (a) Parameter map in the $(p_1,p_2)$ plane for $\sigma=0.2$. The critical line $p_{2,c}=-p_1-\sigma^{-1}$ ($\alpha=\lambda\mathcal G(1)=0$) separates regions of positive and negative KdV quadratic nonlinearity. The diffuse Gardner shading represents an illustrative neighbourhood where quadratic and cubic nonlinearities can be comparable. The vertical section at $p_1=1$ and the three markers identify $\delta=p_2-p_{2,c}=-0.5,0,+0.5$. (b) Signed electrostatic-potential amplitudes of compressive ($A>0$) and rarefactive ($A<0$) solitons along that section at $M-\lambda=0.0012$. Solid curves are admissible exact Sagdeev homoclinic branches, whereas dashed curves are leading Gardner predictions. Filled circles are amplitudes measured in the full time-dependent fP model, initialised with the corresponding exact profiles. Here $\beta_c=8$ ($p_3\simeq35.21$), $p_{2,c}=-6$, and the dotted vertical guide marks $\delta=0$.}\label{fig:curvature_branches}
\end{figure}

The parameter map in Fig.~\ref{fig:curvature_branches}(a) shows the unfolding in the $(p_1,p_2)$ plane. At fixed $\sigma$ and for $\lambda=\sqrt{1+\sigma p_1}>0$, the vanishing quadratic coefficient $\alpha=\lambda\mathcal G(1)$ defines the critical line
\begin{equation}
\label{eq:critical_line_map}
 p_{2,c}(p_1)=-p_1-\frac{1}{\sigma}.
\end{equation}
The critical line of Eq.~\eqref{eq:critical_line_map} corresponds to a codimension-one degeneracy. Above and below this line, $\alpha$ is respectively positive and negative, corresponding to opposite leading-order KdV nonlinearities and soliton polarities for positive excess speed. Exactly on the line, the quadratic coefficient vanishes and the mKdV scaling applies when the cubic term is nonzero. In its vicinity, the quadratic and cubic contributions can become comparable, giving the two-sided Gardner reduction. We remark that the diffuse shading in Fig.~\ref{fig:curvature_branches}(a) is a schematic, amplitude-dependent neighbourhood with no sharply defined phase boundary. The relevant crossover depends on the pulse amplitude as well as the nonlinear coefficients.

The vertical section $p_1=1$ in Fig.~\ref{fig:curvature_branches}(a) identifies the pressure-curvature scan used in Fig.~\ref{fig:curvature_branches}(b), including the three detunings sampled in the direct simulations. To isolate the unfolding of the mKdV degeneracy along this section, we choose the focusing cubic coefficient $\beta_c=8$, which fixes $p_3\simeq35.21$ through Eq.~\eqref{eq:beta_critical}. This choice gives $A_{\rm mKdV}^{\pm}=\pm0.03$ at the selected excess propagation speed $M-\lambda=0.0012$. We then vary $\delta=p_2-p_{2,c}$ about $p_{2,c}=-6$, so that $\alpha=\sigma\delta/\lambda$. Figure~\ref{fig:curvature_branches}(b) compares the exact signed Sagdeev amplitudes with the Gardner predictions obtained from Eq.~\eqref{eq:gardner_speed_physical}. For the latter, solving the quadratic amplitude-speed relation gives
\begin{equation}\label{eq:amplitude_unfolding}
 A_{\rm G}^{\pm}=\frac{-\alpha\pm\sqrt{\alpha^2+6\beta_c(M-\lambda)}}{\beta_c},
 \qquad\beta_c>0.
\end{equation}
At criticality, the two Gardner amplitudes are opposite and equal in magnitude, whereas the exact Sagdeev branches exhibit a small finite-amplitude asymmetry. As $\delta$ changes sign, one branch approaches the conventional KdV polarity while the opposite-polarity branch persists through the near-critical Gardner regime. Thus, the change in curvature reorganises the two soliton families without requiring the disappearance of either branch at $\delta=0$. This coexistence is consistent with the near-critical polarity classification of Olivier and coworkers \cite{OlivierVerheestMaharaj2016}. Figure~\ref{fig:curvature_branches} shows a pressure-curvature parametrisation and a direct comparison of the reduced, exact stationary and time-dependent descriptions.

The resulting two-branch geometry directly illustrates EOS-controlled unfolding. The Gardner approximation follows the exact branches closely near criticality, while finite-amplitude differences become more pronounced with increasing detuning. These departures reflect the local character of the weakly nonlinear reduction. 

\subsection{Changing the sign of the critical cubic nonlinearity}\label{sec:bifurcations_cubic}

A complementary transition occurs along the critical manifold itself. Setting $p_2=p_{2,c}$ and varying $p_3$, Eq.~\eqref{eq:beta_critical} gives
\begin{equation}
\label{eq:cubic_transition}
 \beta_c(p_3)=\frac{3}{2\lambda}(\sigma p_3-\lambda^2),
 \qquad p_{3,c}:=\frac{\lambda^2}{\sigma}=6.
\end{equation}
The simultaneous vanishing of the quadratic and cubic nonlinear coefficients, $\alpha=\beta_c=0$, defines a codimension-two degeneracy in the constitutive parameter space. For the polynomial EOS considered here, this occurs at $p_2=-p_1-\sigma^{-1}$ and $p_3=p_1+\sigma^{-1}$. At this point, both the KdV and mKdV nonlinearities vanish. The next nonlinear coefficient, controlled by the fourth pressure derivative, follows from the multiple-scales hierarchy shown in \S~\ref{sec:nonlinear_hierarchy} and agrees with the exact stationary Sagdeev coefficient in \S~\ref{sec:sagdeev_critical}.

\begin{figure}[tbp]
 \centering
 \includegraphics[width=0.92\linewidth]{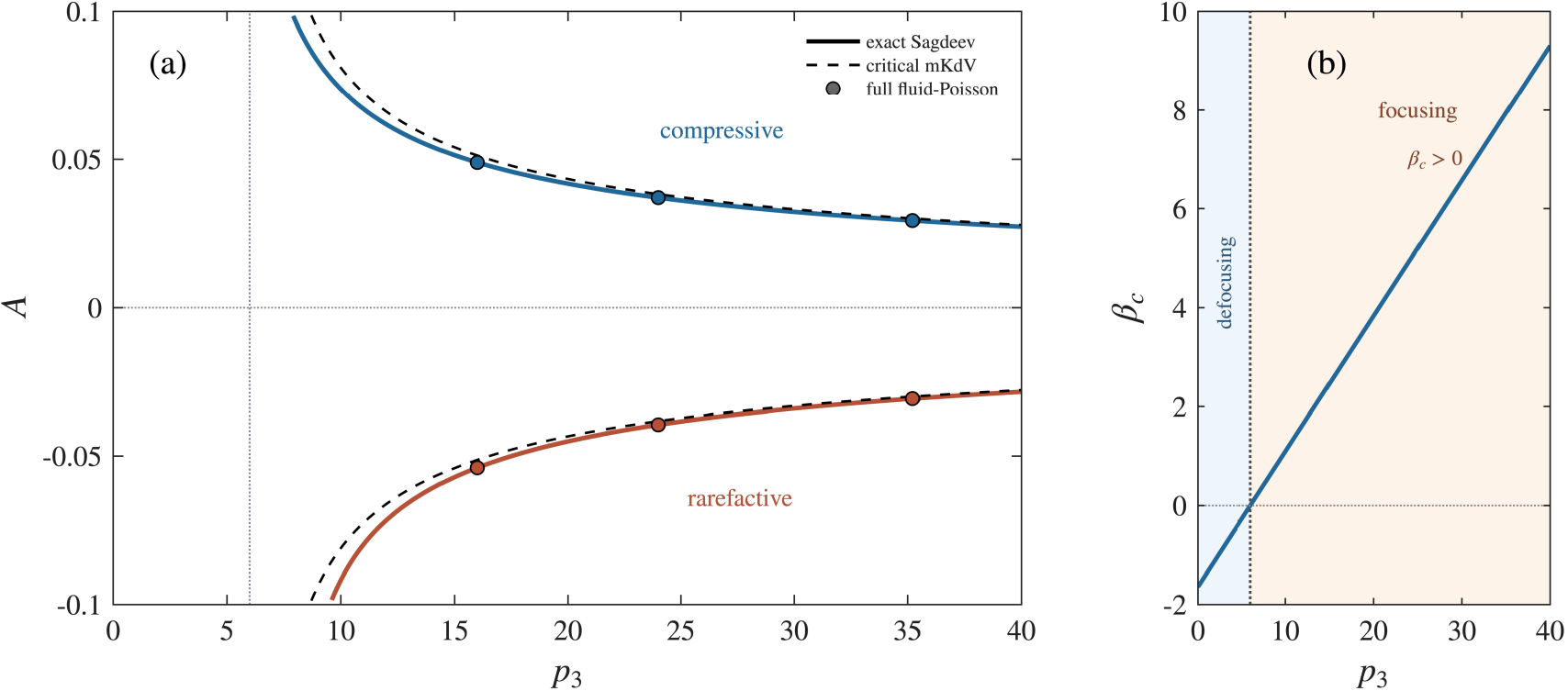}
 \caption{Transition in the cubic critical nonlinearity at $p_2=p_{2,c}=-6$ for $\sigma=0.2$ and $p_1=1$. (a) Signed zero-background pulse amplitudes versus $p_3$ at fixed $M-\lambda=0.0012$. Exact Sagdeev branches (solid), critical mKdV predictions (dashed), and amplitudes measured from full fluid-Poisson evolution initialised with exact Sagdeev profiles (filled circles), shown within a small-amplitude window. (b) Corresponding coefficient $\beta_c(p_3)$, which vanishes at $p_{3,c}=6$. }\label{fig:cubic_transition}
\end{figure}

The sign change at $p_{3,c}$ distinguishes the focusing and defocusing critical reductions. For $\beta_c>0$, mKdV predicts two zero-background pulses with $A_{\rm mKdV}^{\pm}=\pm\sqrt{6(M-\lambda)/\beta_c}$. By contrast, when $\beta_c<0$, this leading-order construction provides no small-amplitude, zero-background bright pulse for $M>\lambda$. Figure~\ref{fig:cubic_transition}(a) compares the compressive and rarefactive soliton amplitudes predicted by the critical mKdV equation with those obtained from the exact Sagdeev formulation and direct numerical simulations of the full fP system. Figure~\ref{fig:cubic_transition}(b) shows the dependence of the cubic nonlinear coefficient $\beta_c$ on $p_3$, highlighting its change of sign at $p_{3,c}=6$, which separates the focusing and defocusing regimes.

The vicinity of $p_{3,c}$ deserves particular care. Because the mKdV amplitude diverges at fixed non-zero speed as $\beta_c\to0^+$, this limit lies outside the assumed weak-amplitude scaling. At $p_{3,c}$ the cubic term also vanishes, and the distinct quartic-nonlinearity scaling of Eq.~\eqref{eq:quartic_evolution} applies when $\lambda^2+\sigma p_4\ne0$. Thus, the sign change in $\beta_c$ defines a local criticality condition, but the apparent termination of a plotted mKdV curve is not, by itself, an exact saddle-node bifurcation of the full model. In the defocusing regime, fronts (double layers) with nonzero asymptotic states must be sought using the appropriate Sagdeev energy level and boundary conditions.

As a finite-time numerical validation of the propagation scheme, we evolved ten initial profiles constructed from the exact Sagdeev pseudopotential. The filled circles in Figs.~\ref{fig:curvature_branches} and \ref{fig:cubic_transition} show their amplitudes measured after propagation under the time-dependent fP equations. For both polarities, these measurements remain indistinguishable from the exact branches on the scale of the figures, and the propagation speeds agree closely with the corresponding travelling-wave predictions. These simulations verify that the selected exact-profile states propagate consistently with their stationary predictions over the simulated interval. Because the observation time is short compared with the slow soliton propagation time, they do not establish long-time persistence, stability under perturbations, or formation from general initial conditions.

\subsection{Spatial equilibria, pitchfork-like geometry and the Gardner fold}
\label{sec:bifurcations_spatial}

\begin{figure}[t]
 \centering
 \includegraphics[width=0.91\linewidth]{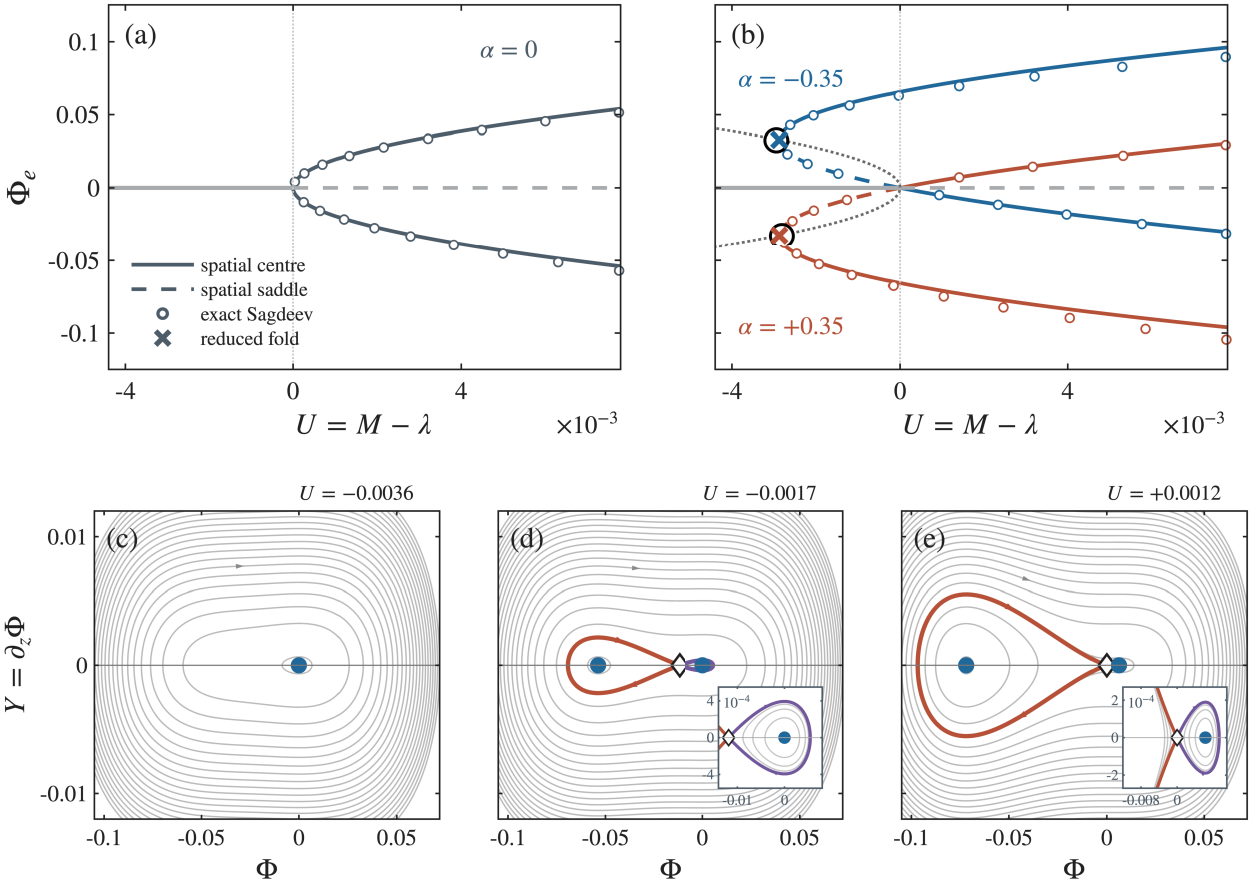}
 \caption{Spatial-equilibrium bifurcations and Hamiltonian phase portraits for $\sigma=0.2$, $p_1=1$, and $\beta_c=8$ ($p_3\simeq35.21$). (a) Pitchfork-like equilibrium structure at $\alpha=0$ and (b) its constrained symmetry-breaking unfolding at $\alpha=\pm0.35$. Solid and dashed curves distinguish spatial centres and saddles of the reduced Gardner system. Open circles indicate exact Sagdeev equilibria, while crosses mark the fold locations. As the value of $\alpha$ changes, the fold location moves along the parabolic dotted curve shown in grey. (c)-(e) Phase portraits of the reduced Gardner travelling-wave system for $\alpha=0.35$ and the indicated values of $U$. Filled blue circles denote spatial centres and white diamonds denote saddles. Copper and purple curves are distinct homoclinic trajectories making negative and positive excursions, respectively, from the saddle background. In (d) the background is nonzero, whereas in (e) it is $\Phi=0$, so the copper and purple loops correspond to rarefactive and compressive zero-background solitons. Insets in (d) and (e) magnify the saddle neighbourhood, displaying the small purple loop and part of the copper separatrix. Thin grey closed curves represent spatially periodic trajectories. Arrowheads show the direction of increasing travelling coordinate $z$.}\label{fig:spatial_equilibria}
\end{figure}

Changes in pulse amplitude can be related to the spatial phase portrait, provided constant travelling-wave equilibria and the extrema of homoclinic trajectories remain distinct. For the physical leading-order Gardner approximation, let $\phi(x,t)\simeq \Phi(z)$,
where $z:=x-(\lambda+U)t$ and $U:=M-\lambda=V$ denotes the physical speed excess. The exact problem uses $\zeta=x-Mt$, while $z_s$ and $U_s$ refer to the critical stretched variables. Thus, the physical counterpart of the scaled Gardner travelling-wave equation~\eqref{eq:gardner_travelling_ode} defines the following planar Hamiltonian system
\begin{equation}
\label{eq:spatial_system}
 \frac{\mbox{d}\Phi}{\mbox{d}z}=Y,\qquad
 \gamma \frac{\mbox{d}Y}{\mbox{d}z}=U\Phi-\frac{\alpha}{2}\Phi^2-\frac{\beta_c}{3}\Phi^3.
\end{equation}
In the focusing regime $\beta_c>0$, let us introduce $\chi:=\sqrt{\beta_c/(3\gamma)}\,\Phi$ and $\Upsilon:=\sqrt{\beta_c/(3\gamma)}\,Y$, together with the control parameters $r:=U/\gamma$ and $\eta:=-(\alpha/2)\sqrt{3/(\gamma\beta_c)}$. At exact criticality ($\eta=0$), the spatial system takes the symmetric pitchfork normal form shown in Fig.~\ref{fig:spatial_equilibria}(a),
\begin{equation}\label{eq:spatial_pitchfork_normal_form}
 \frac{\mbox{d}\chi}{\mbox{d}z}=\Upsilon,\qquad \frac{\mbox{d}\Upsilon}{\mbox{d}z}=r\chi-\chi^3.
\end{equation}
The corresponding equilibrium condition is $\chi_e(r-\chi_e^2)=0$. As $r$ becomes positive, the origin changes from a spatial centre to a saddle, and two symmetric centres emerge at $\chi_e=\pm\sqrt r$. This is the same equilibrium geometry as the familiar scalar supercritical-pitchfork normal form, but the scalar equation is only an auxiliary branch description. The actual travelling-wave dynamics is the two-dimensional Hamiltonian system above. Thus, we remark that  the centre-saddle classification is not directly related to the stability/instability of the plasma.

When the pressure curvature is detuned from criticality ($\eta\ne0$), the governing spatial normal form shown in Fig.~\ref{fig:spatial_equilibria}(b) becomes
\begin{equation}\label{eq:spatial_unfolded_normal_form}
 \frac{\mbox{d}\chi}{\mbox{d}z}=\Upsilon,\qquad \frac{\mbox{d}\Upsilon}{\mbox{d}z}=r\chi+\eta \chi^2-\chi^3.
\end{equation}
The quadratic term breaks the $\chi\mapsto-\chi$ symmetry, making the positive and negative-amplitude equilibrium branches inequivalent. Unlike a generic additive imperfection of a pitchfork, this unfolding retains the reference equilibrium $\chi=0$. In physical variables, its nonzero equilibria satisfy
\begin{equation}\label{eq:spatial_equilibria}
 U=\frac{\alpha}{2}\Phi_e+\frac{\beta_c}{3}\Phi_e^2.
\end{equation}
Their saddle-centre fold occurs at
\begin{equation}\label{eq:spatial_fold}
 \Phi_{e,{\rm fold}}=-\frac{3\alpha}{4\beta_c},\qquad
 U_{\rm fold}=-\frac{3\alpha^2}{16\beta_c}.
\end{equation}
Thus, the fold amplitude approaches zero linearly with $\alpha$, whereas its speed detuning approaches zero quadratically. Equivalently, the normalized fold lies at $\chi_{\rm fold}=\eta/2$ and $r_{\rm fold}=-\eta^2/4$. Figure~\ref{fig:spatial_equilibria} reconstructs this geometry using the Jacobian classification of the two-dimensional spatial system, and compares the reduced equilibrium branches with corresponding equilibria of the exact Sagdeev equation. For the illustrated Gardner detunings, $|\alpha|/\beta_c\leq0.35/8=0.04375<0.1$, satisfying the order-of-magnitude coefficient separation given by Verheest and Hereman \cite{VerheestHereman2025}, although this necessary ordering check does not ensure quantitative accuracy at finite amplitude. For $\alpha=\pm0.35$, the reduced fold is located at $U_{\rm fold}=-0.0028711$. The corresponding exact folds lie near $-0.0029417$ and $-0.0028074$, respectively. These finite-amplitude shifts are consistent with the local character of the Gardner approximation.

The spatial classification also has a direct acoustic interpretation in the parent model. At a uniform equilibrium $(n_\ast,u_\ast,\phi_\ast)$ of the exact travelling-wave problem, $n_\ast=e^{\phi_\ast}$ and $u_\ast=M(1-n_\ast^{-1})$. Consequently, the speed of the background flow relative to the travelling wave is $|u_\ast-M|=M/n_\ast$ for $M>0$. The local long-wave sound speed is
$c(n_\ast)=\sqrt{1+\sigma p'(n_\ast)}$. On a regular branch with $\partial_n\mathcal B(n_\ast;M)>0$, the equilibrium is a spatial saddle
when $M/n_\ast>c(n_\ast)$, and a spatial centre when $M/n_\ast<c(n_\ast)$. Equality marks a sonic, degenerate spatial equilibrium, including the exact saddle-centre fold when the two nonzero branches collide. This fold occurs at negative $U$ for the focusing case $\beta_c>0$, whereas the ordinary zero-background homoclinic pulses bifurcate from the acoustic threshold into $U>0$. It is therefore a fold of spatial equilibria, instead of a fold of the compressive or rarefactive soliton branches displayed in Figs.~\ref{fig:curvature_branches} and \ref{fig:cubic_transition}. 

The lower panels of Fig.~\ref{fig:spatial_equilibria} illustrate the associated reorganisation of the reduced Gardner phase space. Below the fold, Fig.~\ref{fig:spatial_equilibria}(c) contains only the central spatial equilibrium. In Fig.~\ref{fig:spatial_equilibria}(d), a saddle-centre pair has emerged. The two coloured homoclinic trajectories attached to the saddle (white diamond) make negative (copper) and positive (purple) excursions from a nonzero asymptotic potential. They enclose families of periodic trajectories around the corresponding spatial centres (filled blue circles). When $U>0$ [Fig.~\ref{fig:spatial_equilibria}(e)], the saddle moves to the origin, and the two loops represent the rarefactive and compressive soliton branches on the zero background. The insets of Figs.~\ref{fig:spatial_equilibria}(d) and ~\ref{fig:spatial_equilibria}(e) show the small purple homoclinic loops together with nearby portions of the copper separatrices. The distinct orbits meet only at their common saddle.

This conventional spatial-equilibrium analysis helps distinguish the turning points of reduced homoclinic trajectories from the stationary equilibria to which they may connect. Taken together, our results provide a coherent view of EOS-induced dispersive-wave criticality. The vanishing of $\mathcal G(1)$ produces a degeneracy in the right-going acoustic characteristic field. Detuning $p_2$ unfolds the associated focusing soliton branches, while varying $p_3$ through $p_{3,c}$ changes the critical cubic response. The exact travelling-wave problem confirms the admissible homoclinic branches and the spatial-equilibrium folds, while exposing finite-amplitude corrections to the reduced normal forms. The time-dependent simulations provide a complementary check of selected soliton branches without altering this spatial bifurcation interpretation.

\subsection{EOS-controlled double-layer boundary and flat-top solitons}
\label{sec:bifurcations_double_layer}

The transition in the local cubic coefficient at $p_{3,c}=6$ does not, by itself, locate the termination of a finite-amplitude soliton branch. To make this distinction explicit, we continue the exact Sagdeev orbits along $p_2=p_{2,c}=-6$ with $\sigma=0.2$, $p_1=1$ and $V=M-\lambda>0$. A regular zero-background soliton leaves the spatial saddle $\phi=0$, reaches a simple nonzero turning point $A$ satisfying $\mathcal S(A;M)=0$ and $\partial_\phi\mathcal S(A;M)\ne0$, and returns to its initial equilibrium. A double layer arises at the limiting parameter value where that turning point becomes a second spatial equilibrium,
\begin{equation}\label{eq:dl_equilibrium_conditions}
 \mathcal S(A_{\mathrm{DL}};M)=0,\qquad
 \partial_\phi\mathcal S(A_{\mathrm{DL}};M)=0,\qquad
 A_{\mathrm{DL}}\ne0.
\end{equation}
The stationary orbit is then a heteroclinic connection from $\phi=0$ to $\phi=A_{\mathrm{DL}}$. Both equilibria must lie on the same regular Bernoulli branch, with $\mathcal S<0$ between them. The double-root conditions alone are not an existence or temporal-stability criterion. This double-layer mechanism is already established in plasma theory, including the occurrence of nearby flat-top electrostatic solitons \cite{OlivierVerheestMaharaj2016,OlivierVerheestMaharaj2017,VerheestHellbergOlivier2020}.

\begin{figure}[tbp]
 \centering
 \includegraphics[width=0.96\linewidth]{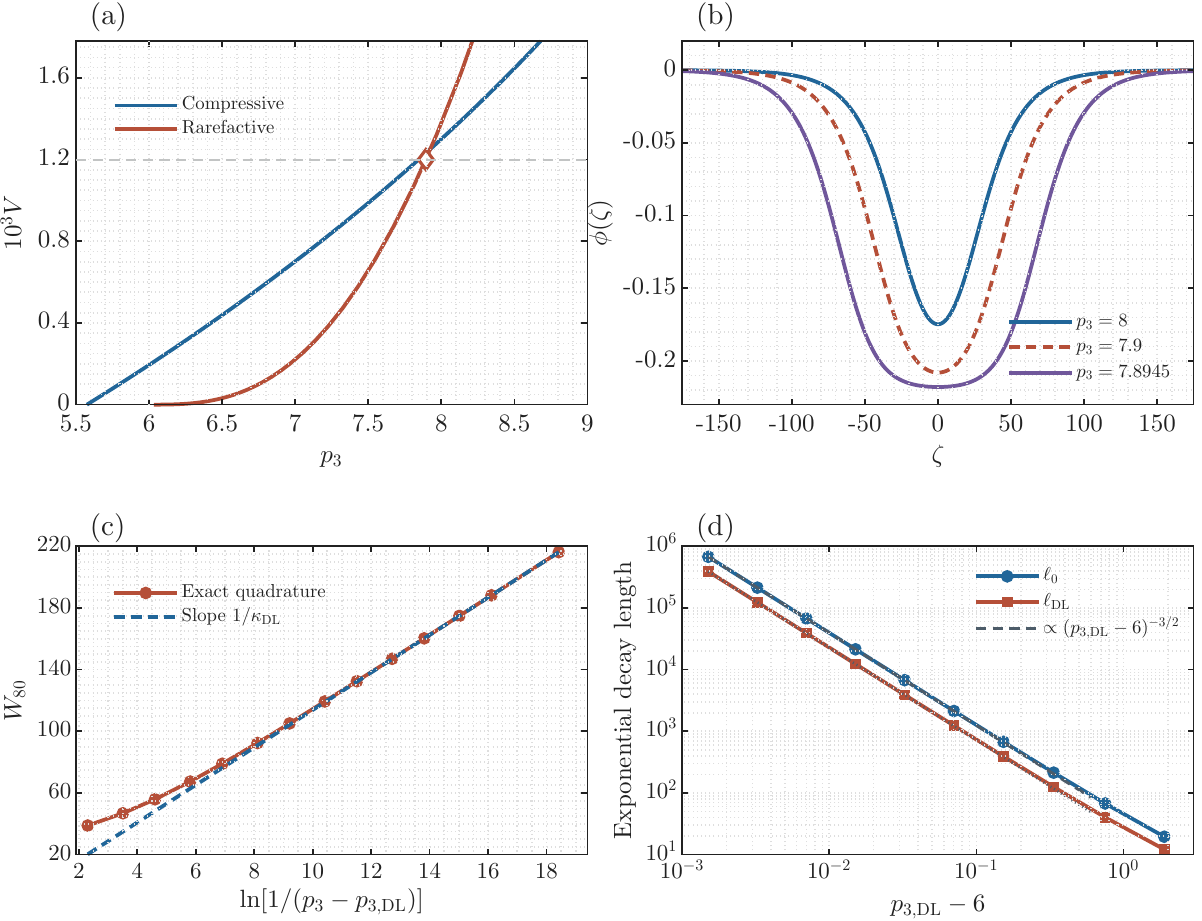}
 \caption{Exact equation-of-state-controlled double-layer boundary and localisation near cubic criticality for $\sigma=0.2$, $p_1=1$ and $p_2=-6$. (a) Rarefactive double-layer locus and the distinct boundary at which the minimum pressure slope sampled by a compressive soliton orbit vanishes. The horizontal line marks $V=0.0012$, and the diamond marks its rarefactive double-layer endpoint. (b) Increasingly flat rarefactive homoclinic profiles at that speed as $p_3$ approaches the double-layer boundary from above. (c) Full $80\%$ plateau width from exact Sagdeev quadrature and the logarithmic asymptotic slope $1/\kappa_{\mathrm{DL}}$. (d) Exponential decay lengths at the two equilibria along the double-layer locus, together with the  $(p_{3,\mathrm{DL}}-6)^{-3/2}$ asymptotics. All curves correspond to stationary calculations.}
 \label{fig:dl_localisation}
\end{figure}

At the reference excess speed $V=0.0012$, the rarefactive homoclinic branch terminates at the exact stationary double-layer boundary $p_{3,\mathrm{DL}}\simeq7.8943955063$, with $n_{\mathrm{DL}}\simeq0.8029337494$ and $A_{\mathrm{DL}}\simeq-0.2194830723$, as shown in 
Fig.~\ref{fig:dl_localisation}(a). This endpoint lies strictly above $p_{3,c}=6$, even though the compressive zero-background branch crosses the cubic critical value at this fixed speed. The exact double-layer locus can be parameterised by a two-by-two linear system for $(M^2,p_3)$ at a chosen endpoint density. Its existence must still be checked along the entire connecting trajectory.

The approach to the rarefactive endpoint has a simple geometrical interpretation. As $p_3$ decreases toward $p_{3,\mathrm{DL}}$ from above, the simple turning point of the homoclinic orbit approaches the nonzero equilibrium $A_{\mathrm{DL}}$. Near this almost-stationary point, the spatial restoring force becomes small, so the potential profile spends an increasingly long distance close to its minimum before returning to the background. This produces the extended plateau shown in Fig.~\ref{fig:dl_localisation}(b). In the limiting case, the return occurs only at infinite distance and the two transitions separate into oppositely oriented double-layer fronts. The soliton itself remains homoclinic for every $p_3>p_{3,\mathrm{DL}}$.

This description also supplies a quantitative law. Let us choose a detuning such that \(\delta = p_3 - p_{3,\mathrm{DL}} > 0\) while keeping \(V = 0.0012\) fixed. Let \(W_{80}\) denote the full separation between the two positions at which the potential attains \(80\%\) of its negative extremal value. A generic splitting of the
nonzero double root gives
\begin{equation}\label{eq:dl_plateau_width}
 W_{80}=\frac{1}{\kappa_{\mathrm{DL}}}\ln\!\frac1\delta+\mathcal{O}(1),
 \qquad
 \kappa_{\mathrm{DL}}^2
 =-\partial_{\phi\phi}\mathcal S(A_{\mathrm{DL}};M)>0.
\end{equation}
This logarithmic divergence is the standard passage-time law near a hyperbolic saddle. The coefficient is
the inverse spatial decay rate at the second saddle and is independent of the fixed fractional threshold used to define the plateau width. In the present example, $\kappa_{\mathrm{DL}}^{-1}\simeq12.15828$. Exact high-precision Sagdeev quadrature gives $W_{80}=188.23957$ and $216.21503$ at $\delta=10^{-7}$ and $10^{-8}$, respectively. The corresponding logarithmic secant slope is
$12.14959$, in close agreement with Eq.~\eqref{eq:dl_plateau_width} and Fig.~\ref{fig:dl_localisation}(c). This agreement is a quantitative consistency check of the exact stationary continuation. These calculations connect the EOS-controlled double-layer boundary, orbitwise admissibility and higher-order criticality of the parent fP system \cite{VerheestHellbergOlivier2020}.

A separate constraint is the pressure slope sampled by the complete orbit. For the homogeneous local fP model, $p'(n)\ge0$ is sufficient
for a real linear frequency at every wavenumber. At $V=0.0012$, the compressive exact branch first satisfies $\min_{\mathrm{orbit}}p'(n)\ge0$ at $p_3\simeq7.844895$, which is distinct from $p_{3,\mathrm{DL}}$. The globally sufficient cubic-EOS bound $p_3\ge12$ is unnecessarily restrictive for this individual profile. In contrast, the rarefactive connecting orbits here lie in $n\le1$, where $p'(n)>0$. Both the Bernoulli regularity condition and the Sagdeev sign condition must also hold throughout the orbit. The pressure-slope threshold is speed-dependent and must not be interpreted as the universal termination of compressive solitons near $p_{3,c}$.

\subsection{Approach to the acoustic codimension-two point}
\label{sec:bifurcations_sonic}

The double-layer locus itself approaches the higher-order constitutive critical point $(p_3,V)=(6,0)$. Define $\Delta=p_{3,\mathrm{DL}}-6>0$. The fifth-order exact Sagdeev expansion, consistent with the quartic-flux evolution hierarchy in Eq.~\eqref{eq:quartic_evolution}, gives the double-layer amplitude and speed
\begin{equation}\label{eq:dl_eos_critical_scaling}
 A_{\mathrm{DL}}=-\frac{5\sigma}{6\lambda^2}\Delta+o(\Delta),
 \qquad
 V_{\mathrm{DL}}=\frac{25\sigma^3}{432\lambda^5}\Delta^3
 +o(\Delta^3).
\end{equation}
These are EOS-specific coefficients for the exact stationary boundary. In particular, along this locus the background localisation length and the length at its second equilibrium satisfy
\begin{equation}\label{eq:dl_acoustic_lengths}
 \ell_0\sim
 \frac{\sqrt{216}\,\lambda^2}{5\sigma^{3/2}}\Delta^{-3/2},
 \qquad
 \frac{\ell_{\mathrm{DL}}}{\ell_0}\longrightarrow\frac1{\sqrt3}.
\end{equation}
Figure~\ref{fig:dl_localisation}(d) compares these asymptotics with exact continuation. The divergence differs from the fixed-speed flat-top limit. There, the plateau grows logarithmically while the outer decay rates stay finite, whereas along Eq.~\eqref{eq:dl_eos_critical_scaling} the decay rates vanish and the double-layer amplitude tends to zero.

Exactly at the acoustic critical point, the local pseudopotential has a fifth-order zero, $\mathcal S(\phi;\lambda)=(\lambda^2/5)\phi^5+O(\phi^6)$. This allows a formal one-sided negative-potential asymptotic tail proportional to $|\zeta|^{-2/3}$, as in the broader theory of algebraic acoustic-speed tails \cite{OlivierVerheestMaharaj2017}. However, the complete Sagdeev orbit does not exist here. An exact sign argument for our cubic EOS excludes a finite-amplitude regular homoclinic or double layer attached to the reference background at $(p_3,V)=(6,0)$. Thus, the fifth-order non-Morse stationary germ controls the singular approach to criticality.

\section{Discussion: Applicability of the constitutive description}
\label{sec:eos_applicability}

The bifurcations identified in this work follow from the local derivatives of a prescribed ion pressure law, independently of how that law is generated microscopically. This generality is useful for isolating the nonlinear-wave mechanism. In particular, the polynomial closure used in Figs.~\ref{fig:curvature_branches} and \ref{fig:cubic_transition} is a controlled constitutive model. Its purpose is to vary the effective acoustic nonlinearity without simultaneously changing the linear restoring force. We therefore distinguish the mathematical accessibility of the critical manifold from the thermodynamic and kinetic conditions required for its physical realisation.

\subsection{Acoustic stability and a correlation-pressure benchmark}\label{sec:eos_stability}

For any smooth barotropic closure, the necessary long-wave condition is $\lambda^2=1+\sigma p'(1)>0$. In contrast, the exact dispersion relation of the local fP system, Eq.~\eqref{eq:dispersion_exact}, imposes a stronger requirement if stability is demanded over its entire Fourier spectrum,
\begin{equation}
\label{eq:eos_full_spectrum}
 \frac{\omega^2(k)}{k^2}=\sigma p'(1)+\frac{1}{1+k^2}\geq0.
\end{equation}
For $\sigma>0$, this holds for every finite $k$ when $p'(1)\geq0$. Thus a real long-wave acoustic speed does not, by itself, exclude an instability at shorter wavelengths. The linear-dispersion tests and exact-Sagdeev branch-marker simulations use $p'(1)=1$ and satisfy this equilibrium linear-stability condition. The separate parameter sweeps of Gaussian-fission solitons must be assessed for their individual EOS settings. Importantly, the criticality condition $\mathcal G(1)=0$ constrains the pressure curvature, not necessarily the sign of its first derivative, and is therefore not intrinsically incompatible with Eq.~\eqref{eq:eos_full_spectrum}.

The conditions on a nonuniform profile are more restrictive than the reality of the sound speed at $n=1$. Around any homogeneous equilibrium density $n_\ast>0$, the unregularised local fP model has the Doppler-shifted linear dispersion relation
\begin{equation}\label{eq:eos_local_dispersion}
(\omega-ku_\ast)^2=k^2\left[\sigma p'(n_\ast)
+\frac{n_\ast}{n_\ast+k^2}\right].
\end{equation}
Thus, $c^2(n_\ast)=1+\sigma p'(n_\ast)>0$ ensures a real long-wave sound speed, but stability of all Fourier wavenumbers in this local model requires the stronger condition $p'(n_\ast)\geq0$. For a heterogeneous Sagdeev profile, it is useful to inspect $p'(n)$ over the entire density interval sampled by the orbit. For the cubic closure on the quadratic critical manifold, $p'(n)=p_1+2p_{2,c}(n-1)+3p_3(n-1)^2$. When
$p_1>0$ and $p_3>0$, a sufficient condition for this derivative to be nonnegative at every density is
\begin{equation}\label{eq:eos_global_slope_bound}
p_3\geq\frac{p_{2,c}^2}{3p_1}
=\frac{\lambda^4}{3\sigma^2p_1}.
\end{equation}
For $\sigma=0.2$ and $p_1=1$, the bound is $p_3\geq12$, or equivalently $\beta_c\gtrsim 1.643$. More generally, when $p_1>0$ and $0<\sigma p_1<1/2$, this global nonnegative-pressure-slope condition on the critical cubic EOS implies $p_3\geq\lambda^4/(3\sigma^2p_1)>\lambda^2/\sigma=p_{3,c}$, and hence $\beta_c>0$. Such an EOS must have a focusing critical mKdV reduction. The statement assumes nonnegative $p'(n)$ for all positive densities, which is stronger than hyperbolicity of the quasineutral long-wave system alone. The numerical examples at $p_3=16$, $24$, and $35.2118697$ satisfy it, so does the entire $p_2$ detuning scan at the last $p_3$. Failure of this sufficient global bound does not, however, imply that a particular finite-amplitude orbit crosses a region with $p'(n)<0$. Such a claim requires an interval-specific test. Nor does satisfying the slope bound by itself establish thermodynamic or kinetic realisability of the polynomial EOS.

As a concrete comparison, consider the classical one-component plasma (OCP), whose strongly coupled pressure is represented by Monte Carlo-informed thermodynamic approximations \cite{KhrapakKhrapak2021OCP}. Along an isothermal density variation, a commonly used dimensionless pressure fit gives
\begin{equation}\label{eq:eos_ocp_benchmark}
 p_{\mathrm{OCP}}(n)=0.9071n-0.3\Gamma_0 n^{4/3}
 +0.1981\Gamma_0^{1/3}n^{10/9},
\end{equation}
where $\Gamma_0$ is the equilibrium ionic Coulomb coupling parameter and $\Gamma(n)=\Gamma_0 n^{1/3}$. We use this expression only to probe the accessibility of the derivative condition. For $\Gamma_0=20$, imposing $\mathcal G(1)=0$ formally yields $\sigma_c\simeq0.1283$, $\lambda^2\simeq0.1668$ and $\beta_c\simeq-0.478$. Thus, the correlation pressure can produce the required quadratic degeneracy and a defocusing cubic reduction while retaining a real long-wave sound speed. However, $p'_{\mathrm{OCP}}(1)\simeq-6.495$, so Eq.~\eqref{eq:eos_full_spectrum} predicts $\omega^2(k)<0$ above $k\simeq0.447$ in the unregularised local model. Moreover, for singly charged ions at equal equilibrium electron and ion densities, the associated electron coupling parameter is $\Gamma_e=\sigma_c\Gamma_0\simeq2.57$, inconsistent with a weakly coupled classical Boltzmann-electron population. These observations rule out this example as a self-consistent physical realisation under the present assumptions and are unrelated to the short-wave instability of the actual OCP, whose correlations need not be adequately described by a local pressure term at such scales. Its defocusing value of $\beta_c$ does not contradict the focusing corollary above, since $p'_{\mathrm{OCP}}(1)<0$ and thus the benchmark does not satisfy the required nonnegative pressure-slope hypothesis.

\subsection{Screening and thermodynamic paths}
\label{sec:eos_screening}

A more realistic treatment of correlated ions generally requires electron screening and an EOS that distinguishes particle correlations from the response of the neutralising medium. Thermodynamic descriptions of Yukawa fluids provide useful pressure and sound-speed estimates \cite{KhrapakThomas2015EOS, KhrapakThomas2015Sound}, but their application here is not direct. The screening parameter and the ion coupling change with density along a prescribed isothermal or adiabatic path, while an independently included Boltzmann-electron force and Poisson equation already account for part of the mean-field response. Adding a total screened-plasma pressure without identifying its constituent terms could therefore count the same electrostatic contribution twice. A quantitatively predictive plasma application must derive the appropriate residual ion pressure from a consistent thermodynamic model and verify electron classicality, weak coupling, and the wavelength range of validity of the local fluid approximation.

The present results consequently establish a constitutive mechanism rather than an experimental material assignment. If a physically admissible pressure law reaches $\mathcal G(1)=0$ with a stable acoustic branch, then its first non-vanishing higher derivative determines the local nonlinear-wave class, while the exact Sagdeev formulation specifies the accessible travelling-wave structures. The existence and accuracy of these structures in our numerical experiments refer to the stated fP equations with the chosen polynomial closure. A screening-consistent correlation pressure, possible kinetic or viscoelastic corrections, and the identification of experimentally accessible critical states remain natural extensions, without altering the analytical connection between the fundamental derivative and the KdV-Gardner criticality demonstrated here.

\section{Conclusions}
\label{sec:conclusions}

In this work, we investigated how the ion EOS controls the nonlinear propagation of ion-acoustic waves in a one-dimensional fP plasma. Our analysis shows that the transition between KdV, mKdV and Gardner dynamics is governed by the density dependence of the effective acoustic speed. In the quasineutral limit, this dependence is characterised by the fundamental derivative $\mathcal G$ of the effective barotropic pressure. We obtained the relations $\alpha=\lambda\mathcal G(1)$ and, on the critical manifold, $\beta_c=\lambda\mathcal G'(1)/2$. Thus, the disappearance of quadratic nonlinearity corresponds to the loss of genuine nonlinearity of the acoustic characteristic field, while the next density derivative determines the focusing or defocusing character of the cubic reduction. We further establish both a formal leading-order evolution hierarchy on successive EOS critical manifolds and an exact corresponding hierarchy of Sagdeev derivatives. At the codimension-two point, the full fP reduction yields a quartic-flux equation whose coefficient, profile and leading speed-amplitude relation agree with the independent fifth-order Sagdeev expansion. This builds on established links between constitutive nonlinearity and dispersive-fluid reductions, and specifies them for the present fP closure and its exact Sagdeev hierarchy.

We connect our local description to the exact travelling-wave formulation of the parent system. The Sagdeev pseudopotential provides an independent characterisation of the critical point and identifies which roots of the reduced amplitude-speed relations correspond to admissible solitons. By varying the pressure curvature across its critical value, we found a two-sided Gardner unfolding of compressive and rarefactive homoclinic branches. Variation of the next pressure coefficient changes the sign of the critical cubic nonlinearity and, consequently, the class of small-amplitude structures supported by the reduced equations. The associated spatial equilibria exhibit a pitchfork-like geometry at criticality and saddle-centre folds away from it. On the regular exact branch, these correspond to changes between subsonic and supersonic backgrounds in the travelling frame. Importantly, these equilibrium bifurcations are distinct from the branches of soliton maxima and do not, by themselves, establish temporal stability.

The exact rarefactive branch also terminates at a double layer whose location is controlled by the cubic EOS coefficient. The approach generates logarithmically widening flat-top homoclinics even though both outer tails remain exponential at fixed speed. The logarithmic law itself is a standard saddle-passage effect, here used to validate the model-specific stationary continuation. The double-layer boundary differs from the orbitwise pressure-slope threshold and approaches the codimension-two acoustic degeneracy according to a cubic speed-detuning law. In this latter limit, the amplitude vanishes, and the decay lengths diverge. The formal algebraic tail at the exactly degenerate point does not belong to a complete finite-amplitude regular soliton. These quantitative results specify an EOS-controlled organisation of nonlinear coherent structures that are important for plasma dynamics.

Direct simulations of the full fP equations support this picture. We observed a broad initial disturbance splitting into a train of localised waves and independently verified the linear dispersion relation. Moreover, solitons initialised from exact Sagdeev profiles retained their shape and propagation speed over the simulated intervals. The measured amplitudes for selected compressive and rarefactive states lie on the exact branches as we vary the quadratic and cubic pressure coefficients. These results establish accurate finite-time numerical propagation of the constructed exact-profile states within the prescribed model.

Finally, we formulate the criticality mechanism for a general smooth barotropic closure, where we used a polynomial EOS to resolve the bifurcation structure in a controlled constitutive example. Our comparison with a correlation-pressure benchmark illustrates that formal criticality need not coincide with a physically admissible state once electron coupling and the full linear spectrum are taken into account. Identifying a screening-consistent EOS that reaches this regime is therefore an interesting extension. More broadly, the link between the fundamental derivative, dispersive normal forms and exact travelling-wave geometry provides a framework for understanding how constitutive nonlinearities reorganise coherent structures in dispersive fluids.

\begin{acknowledgments}
The authors thank the support of the Competition for Research Regular Projects, year 2023, code LPR23-06, Universidad Tecnol\'ogica Metropolitana. This work was partially funded by the Agencia Nacional de Investigación y Desarrollo (ANID—Chile) through the FONDECYT de Iniciación grant 11251397 (JFM).
\end{acknowledgments}

\bibliographystyle{apsrev4-2}
\bibliography{references}
\end{document}